\documentclass[twocolumn]{pazhb_eng}

\usepackage{graphicx}
\usepackage{amsmath}
\usepackage{rotating}
\usepackage{lscape}
\usepackage{ifpdf}

\newcommand {\be}{\begin{equation}}
\newcommand {\ee}{\end{equation}}

\def\ergs{erg s$^{-1}$}

\begin{document}
\journalinfo{2015}{41}{8}{394}[406]

\title{IGR\,J17463-2854, a Possible Symbiotic Binary System in the Galactic Center Region }
\author{\bf D.I. Karasev\email{dkarasev@iki.rssi.ru}\address{1},
S.S. Tsygankov\address{2}, A.A.Lutovinov\address{1}\\
\it{$^1$ Space Research Institute, Moscow, Russia\\
$^2$ Tuorla Observatory, University of Turku, Finland}}
\shortauthor{}
\shorttitle{}
\submitted{3 March 2015}

\begin{abstract}
This paper is devoted to determining the nature of the hard X-ray source IGR\,J17463-2854
located toward the Galactic bulge. Using data from the INTEGRAL and Chandra X-ray observatories, we
show that five point X-ray sources with approximately identical fluxes in the 2–10 keV energy band
are detected in the error circle of the object under study. In addition, significant absorption at low energies
has been detected in the spectra of all these sources. Based on data from the VVV (VISTA/ESO) infrared
Galactic Bulge Survey, we have unambiguously identified three of the five sources, determined the $J, H$
and $K$ magnitudes of the corresponding stars, and obtained upper limits on the fluxes for the remaining
two sources. Analysis of the color–magnitude diagrams has shown that one of these objects most likely
belongs to a class of rarely encountered objects, symbiotic binary systems (several tens are known with
certainty), i.e., low-mass binary systems consisting of a white dwarf and a red giant. Note
that all our results were obtained using improved absorption values and an extinction law differing in this
direction from the standard one.

\medskip
\end{abstract}

\section*{INTRODUCTION}

The X-ray source IGR\,J17463-2854 was discovered by the INTEGRAL observatory in the survey
of the Galactic center region (Bird et al. 2010) and up to now this work has been the only one
in which this object was mentioned. The authors determined the coordinates of the source,
 R.A.= 17$^{\rm h}$46$^{\rm m}$21$^{\rm s}$ и Dec.= -28$^\circ$54\arcmin25\arcsec, and provided Xray
flux estimates in the 20–40 and 40–100 keV energy bands:  $\approx5$ and $\approx3.1$ mCrab, respectively.
 At the same time, they pointed out that one should not rely on these estimates due to the noticeable “confusion”
effect  (overlapping of nearby sources),that arises when very crowded sky regions (for example, the Galactic center) are observed by instruments with an insufficient angular resolution. Bird et al. (2010)
also point out that the source is transient in nature, but they did not provide its light
curves or the time intervals where IGR\,J17463-2854 was detected in a high state.

Until recently, the lack of the available information and the complexity and crowdedness of the Galactic
center region have made it impossible not only to determine the nature of this source but also to confirm
its existence in the first place. In this paper, we have performed for the first time
a detailed analysis of the available observational data in different wavelength ranges to determine the
emission characteristics and to establish the nature of this object. It is the next one in our series of
papers devoted to determining the nature of X-ray sources from the ASCA and INTEGRAL Galactic
center surveys (Lutovinov et al. 2015).

\section*{INSTRUMENTS AND DATA ANALYSIS}

\begin{figure*}
\centering
\vbox{
\includegraphics[width=0.9\textwidth,bb=0 0 540 234,clip]{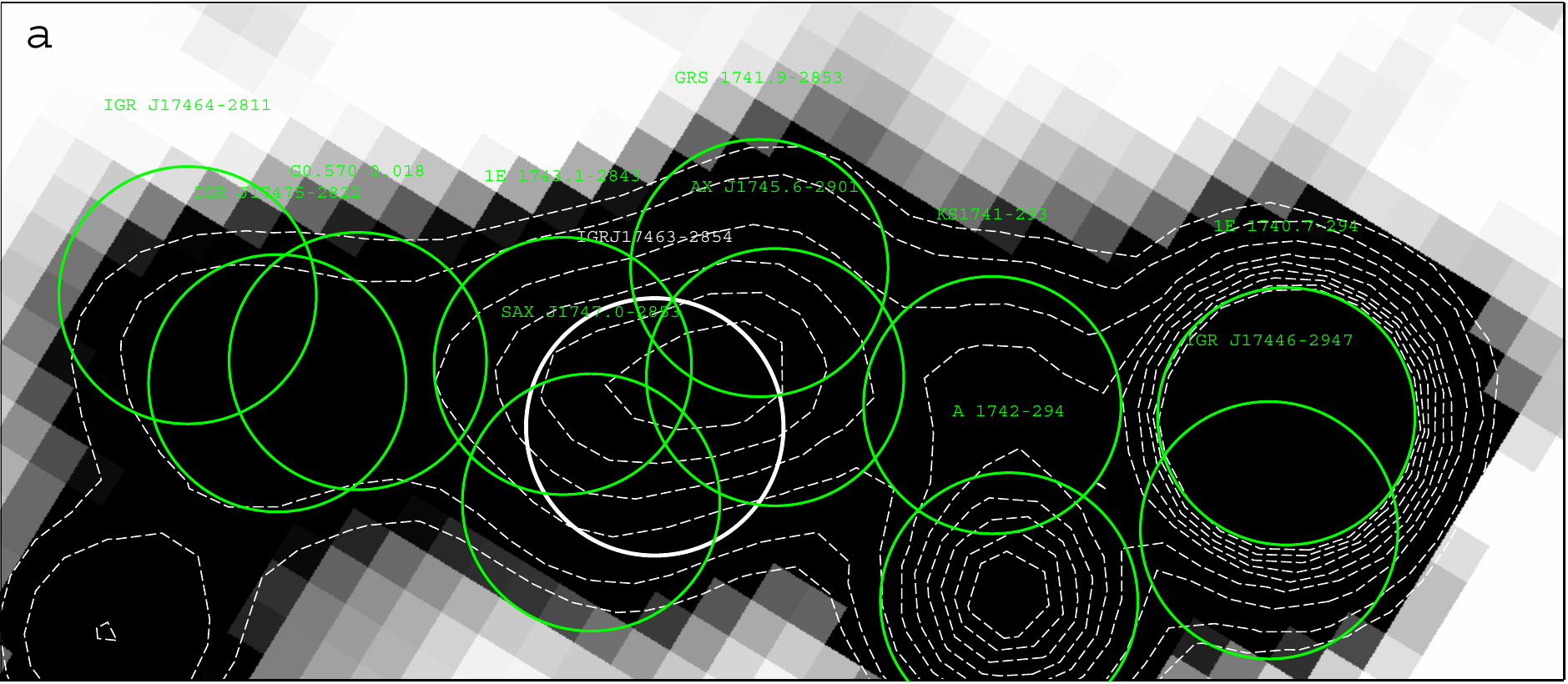}
\includegraphics[width=0.9\textwidth,bb=0 0 540 257,clip]{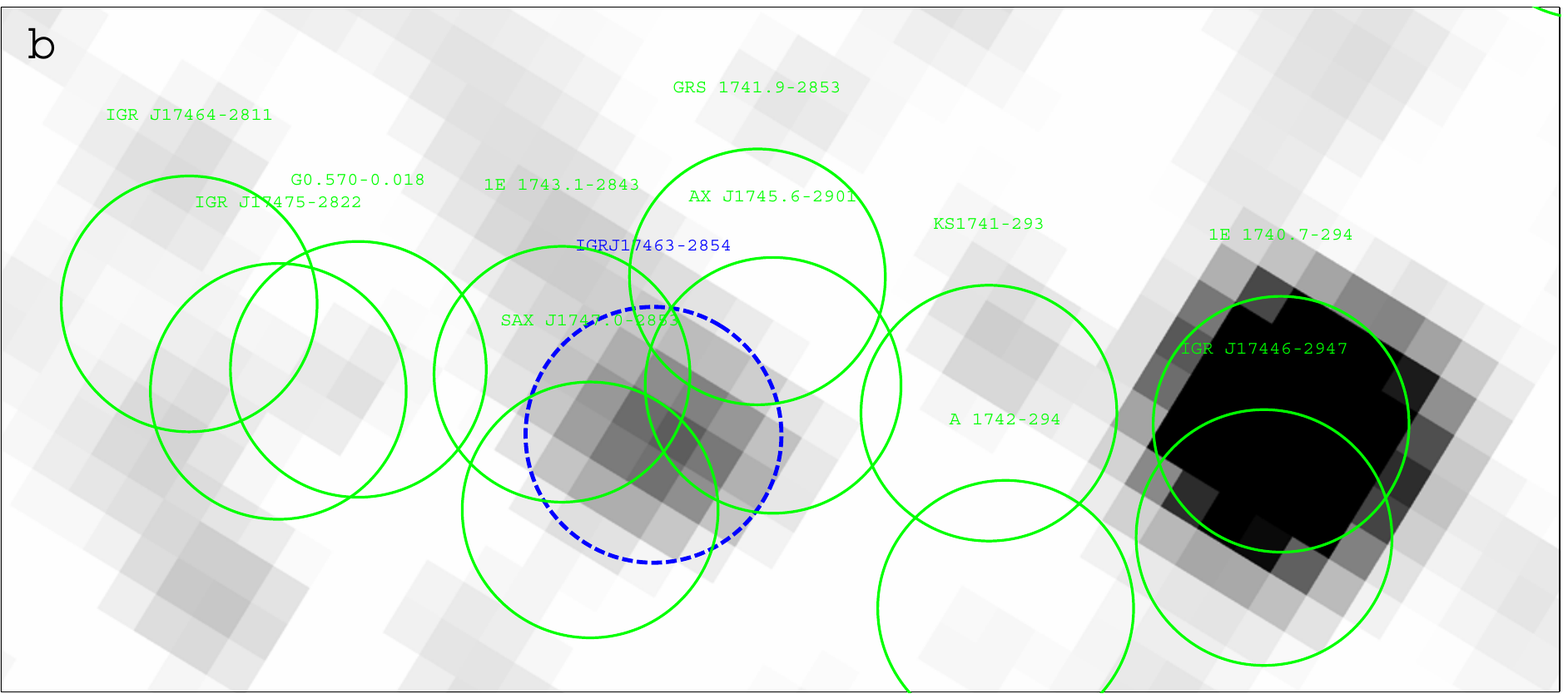}
\includegraphics[width=0.9\textwidth,bb=0 0 540 258,clip]{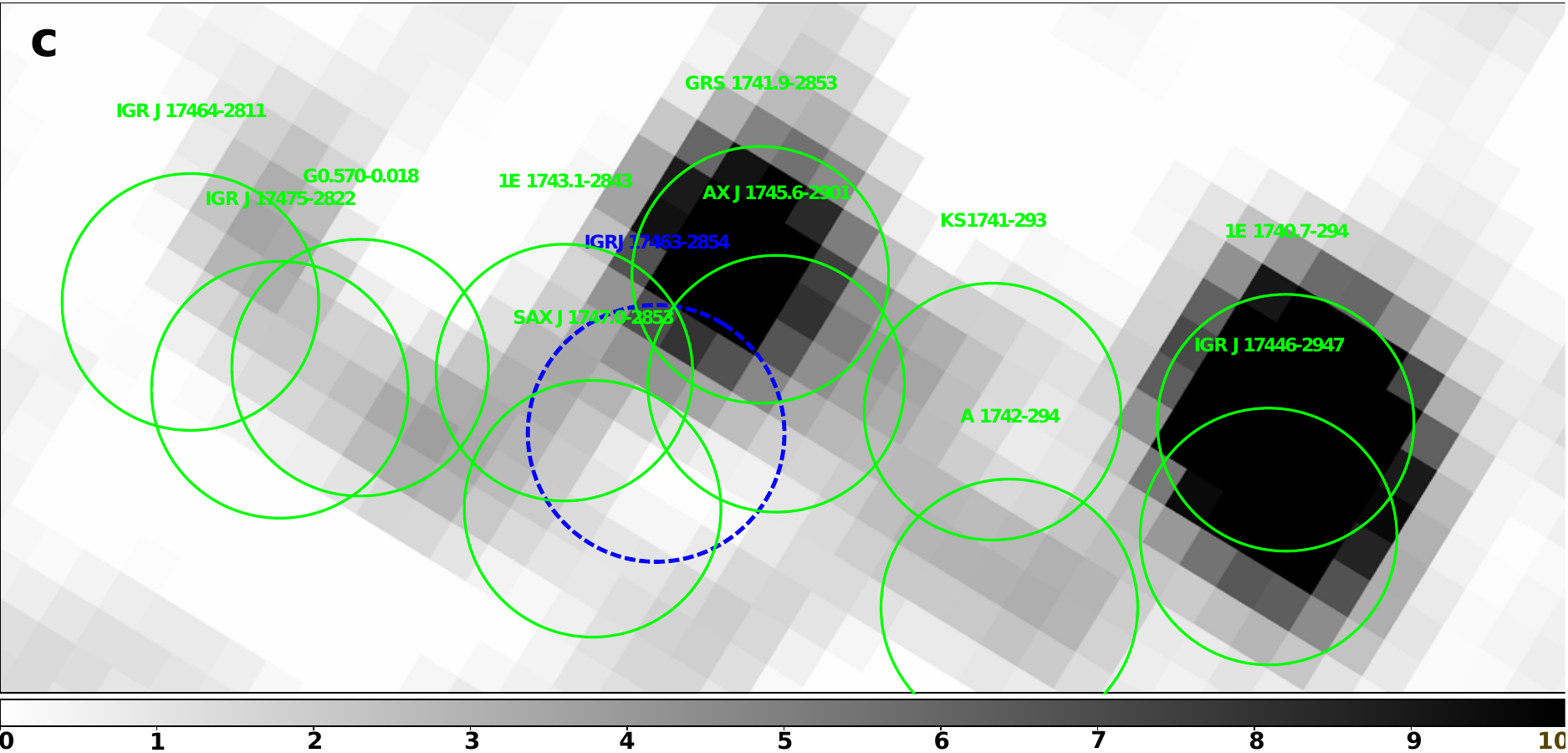}
}
\caption{\footnotesize{ INTEGRAL images of the sky($3^\circ \times1^\circ$) around IGR\,J17463-2854: (a) the image
averaged over the entire time of observations (about 12 years), the outer contour corresponds to a significance
level of $10\sigma$, the remaining ones correspond to those from 16.67 to $133.33\sigma$ with a step of
$16.67\sigma$; (b) the image averaged over orbits 667, 844, and 1025 in which the object under study was detected
with the highest significance; (c) the image obtained for orbit 1201, when IGR\,J17463-2854 was not detected at a
statistically significant level. The image color coding reflects the detection significance and is given in
standard deviations. The circle radius is 12\arcmin, corresponding to the angular resolution of the \textit{IBIS}
telescope.}}  \label{IGRima}
\end{figure*}

To determine the characteristics of IGR J17463-2854 in hard X rays, we used data from the \textit{ISGRI}
detector of the \textit{IBIS} telescope (Ubertini et al. 2003) onboard the INTEGRAL observatory (Winkler et al.
2003) sensitive to photons with energies above 20 keV. The data from this instrument were processed
and analyzed in accordance with the methods and algorithms described in Krivonos et al. (2010) and
Churazov et al. (2014). It is important to note that in this paper we used the data over all 12 years
of INTEGRAL observations, which allowed us to obtain deep maps of the Galactic center region and
to investigate the behavior of the source on various time scales.

To localize the object under study more accurately and to determine its characteristics in the soft X-ray
band, we used Chandra data, the observations with ObsID 945 (July 7, 2000) and 14 - 897 (August 7,
2013), in which the center of the observatory's field of view was closest to the source localization center in
the hard X-ray band. These data were processed using the {\sc CIAO-4.7}, software package and the CALDB
v4.6.5\footnote{http://cxc.harvard.edu/ciao/}. The X-ray spectra of the possible counterparts
to IGR J17463-2854 were fitted using the {\sc XSPEC v12.7} package.

To identify the object optically and to study its properties in the infrared, we used the
\emph{VVV}/ESO (http://www.eso.org/sci/observing/phase3/data\_ releases.html), survey,
which data for the corresponding sky field were reprocessed using the methods of PSF photometry
and the {\sc DAOPHOT-II}, procedures from the {\sc SCISOFT}\footnote{http://www.eso.org/sci/software/scisoft/}
software package. It was done to obtain more accurate photometric magnitudes of stars
in the crowded Galactic center region. To obtain the photometric solutions, we used
2MASS\footnote{http://www.ipac.caltech.edu/2mass/} as a reference catalog. To check the results
of our photometric analysis for correctness, we used data from the
UKIDSS/GPS\footnote{http://surveys.roe.ac.uk/wsa/} catalog.

\begin{figure}
\includegraphics[width=\columnwidth,bb=0 0 534 256,clip]{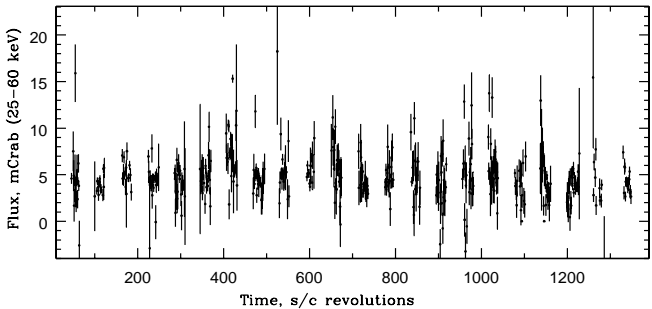}
\caption{\footnotesize{\textit{IBIS/INTEGRAL} light curve from the error circle of IGR\,J17463-2854
in the 25--60 keV energy band. The time is specified in units of the INTEGRAL orbit ($\approx3$ days);
the beginning of the first orbit corresponds to the INTEGRAL launch date, October 17, 2002
(MJD 52564).}}\label{int_lcurve}
\end{figure}

\section*{THE HARD X-RAY BAND. INTEGRAL RESULTS}

The high density of X-ray sources in the Galactic center region and the insufficiently high angular
resolution of the \textit{IBIS/INTEGRAL} telescope (about 12 arcmin) do not allow the total
flux to be separated with confidence between individual objects in the image averaged over
the entire time of observations. This fact is illustrated by Fig. 1a, which shows the image of
a sky field around the Galactic center in the 25--60 keV energy band obtained over $\simeq12$
of INTEGRAL operation. In particular, it can be seen from this figure that there are
four known hard X-ray sources in close proximity to the presumed position of the object
under study:  -- GRS\,1741.9-2853, 1E\,1743.1-2843, AX\,J1745.6-2901 and SAX\,J1747.0-2853.

Nevertheless, owing to the transient nature of all these objects, it turned out to be possible to find
the time intervals when IGR\,J17463-2854 was detected with a higher significance than any of them.
As an example, Fig. 1b presents an image of the same sky field as that in Fig. 1a but averaged over
three orbits of the observatory (667, 844, and 1025) in which the detection significance of IGR\,J17463-2854
was at least twice that for other sources. It can be seen that the localization center of the object
in this image coincides with the position from Bird et al. (2010) and that its flux during these observations
is not affected significantly by neighboring sources and is $10.2\pm1.1$ mCrab in the 25--60 keV
energy band (1 mCrab$\simeq9.44\times10^{-12}$ erg s$^{-1}$ cm$^{-2}$ in this band). The light curve
obtained from the error circle of IGR\,J174632854 (Fig. 2) shows that this flux is close to the
maximum one for the object under study (it should be noted that because of the insufficient angular
resolution of the \textit{IBIS} telescope and the
influence of neighboring sources, the presented curve is not the light curve of IGR\,J17463-2854 in the
ordinary sense; it more likely reflects the maximum possible fluxes for it at each instant of time).

In the INTEGRAL data, there are also periods when IGR\,J17463-2854 was not detected at a statistically
significant level. An example of such a state of the source is presented in Fig. 1c, which shows
the same sky field as that in the previous figures but for orbit 1201. It can be clearly seen that another
source, GRS\,1741.9-2853, was detected at a statistically significant level in this period, while only an
upper limit of 2.6 mCrab (3$\sigma$, 25--60 keV) can be obtained for the flux from IGR\,J17463-2854. The
noticeable and statistically significant increase in flux observed near the four hundredths orbits is associated
with a powerful flare from the transient source SAX\,J1747.0-2853 located in the immediate vicinity
of IGR\,J17463-2854 and affecting the measured flux from it (Fig. 2).
Thus, using all of the available INTEGRAL data, we can confirm the existence of a variable hard X-ray
source with coordinates coincident with those from Bird et al. (2010) and a localization accuracy of $\simeq2.5\arcmin$ (Krivonos et al. 2010).

\begin{figure*}[htb]
\centering
\includegraphics[width=0.9\textwidth,bb=0 0 513 289,clip]{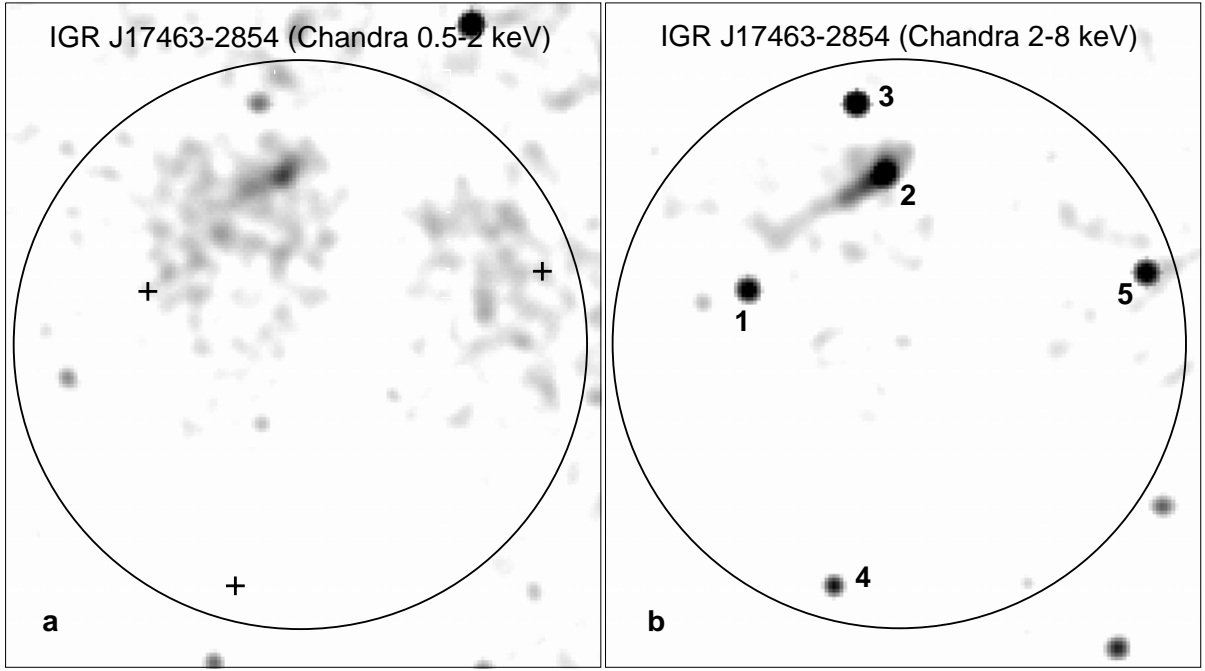}
\caption{\footnotesize{{\it ACIS}/Chandra image of the sky around IGR\,J17463-2854 in two energy
bands, 0.5--2 (left) and 2--8 keV (right). The large circle is the INTEGRAL error circle of
IGR\,J17463-2854. The source numbers reflect their distance from the localization
center of IGR\,J17463-2854. The crosses on the left panel indicate the positions of the undetectable
sources 1, 4, and 5.}}  \label{ChIM}
\end{figure*}

\section*{THE SOFT X-RAY BAND. CHANDRA RESULTS}

Once we have confirmed that IGR\,J17463-2854 really exists in hard X-rays and determined its localization
accuracy, we can search for and investigate its properties in the soft X-ray energy band. The
Chandra X-ray observatory with a high angular resolution is the most suitable instrument for solving
this problem. We searched for sources in the Chandra field of view and determined their astrometric
positions using the {\sc celldetect/CIAO} procedure. Thus, we detected five point and one
extended objects within the INTEGRAL error circle of IGR\,J17463-2854 at a statistically significant level
(Fig. 3). It should be noted that the point objects were known previously and were cataloged in
Evans et al. (2010). However, since the sources were closest to the center of the {\it ACIS} field
of view in the observations we used, we managed to localize them with a better accuracy
($\simeq0.7-1$\arcsec) than that in the above paper. The corresponding coordinates of the sources
are given in Table 1.

Measuring the emission characteristics of the objects detected by Chandra and identifying them with
IGR\,J17463-2854 (searching for the so-called soft X-ray counterpart) were the next step of our studies.
Since one of the criteria for the selection of a counterpart among the possible candidates is the hardness
of its spectrum, first of all we constructed an image of the error circle of IGR\,J17463-2854 in two energy
bands: 0.5--2 and 2--8 keV (Fig. 3). It can be seen from the figure that all five objects are well detected in
the hard channels, with the soft emission from three of them being absent in the 0.5--2 keV energy band,
suggesting significant absorption of their emission. The signal from the remaining two sources
(nos. 2 and 3) is also registered, but its intensity is considerably lower than that in the 2--8 keV
energy band, which also suggests the presence of absorption in the spectra
of these sources. In this paper, we used the Chandra data obtained during two pointings spaced
more than 10 years apart. Since our analysis revealed no statistically significant changes in the parameters
of the spectra for the sources under study, below we summed these observations to improve the statistics.

The spectra of all five sources detected by Chandra are shown in Fig. 4, while their best-fit parameters are
given in Table. 1. We used a simple power law with photoabsorption at low energies to fit the spectra.
As has been assumed above, all sources are strongly absorbed and have similar (given the measurement
errors) slopes and fluxes. Although the absorption estimates obtained by fitting the X-ray spectra of all
sources have significant uncertainties, they show a considerable excess of the hydrogen column density
(at solar heavy-element abundances) above the value in standard catalogs for this sky field,
$N_H\sim1.15\times10^{22}$ cm$^{-2}$ (Kalberla et al. 2005).

It should be noted that we cannot unambiguously assert that intrinsic absorption is present in all sources, because the angular resolution of the neutral
hydrogen radio maps (approximately half a degree) is too low to fully take into account all peculiarities
of the interstellar medium, especially in such a complex region as the Galactic center. However, we
can try to estimate $N_H$ using the infrared extinction data from Gonzalez et al. (2012) obtained with an
angular resolution up to 2\arcmin\ and the known correlation of the gas and dust distributions in the Galaxy,
$A_V = 5.2\times10^{-22} N_H$ (Bohlin et al. 1978).
The extinction in the V band ($A_V$) can be estimated by recalculating
it from the extinction $A_{Ks}$ in the Ks band (Gonzalez et al. 2012) by assuming that the standard extinction
law is valid everywhere (Cardelli et al. 1989; Schlegel et al. 1998). Thus, we can estimate the hydrogen
column density to the Galactic center, $N_{H,ir}\simeq5\times10^{22}$ cm$^{-2}$, which is comparable, within the error
limits, to the derived values for some sources (see Table 1).

\hspace{-0.5cm}\begin{table*}
\centering
\footnotesize{
\begin{minipage}{200mm}
\caption{\footnotesize Characteristics of the possible soft X-ray counterparts to IGR\,J17463-2854}
\begin{tabular}{@{}lccccrrc@{}}
\hline
 \#& Name 		& R.A.(J2000)  		   & Dec (J2000) 				&  $\Gamma$ 	    &  $N_H$, 		& \multicolumn{2}{c}{Flux, 2-10 KeV,  $10^{-13}$ erg/s/cm$^{2}$} \\
   & Evans et al. (2010)& 		  			   & 							&  					&  $10^{22}$ cm$^{-2}$	&  measured   &	     unabsorbed   \\
 \hline
 1 &CXO\,J174627.0-285356 & 17$^h$46$^m$27$^s$.022 & -28$^\circ$53\arcmin56\arcsec.28 & 0.65$_{-0.64}^{+0.74}$ & 16.5$_{-5.3}^{+6.5}$	&
 $2.15_{-0.64}^{+0.74}$		& $3.58$	\\
 2 &CXO\,J174621.6-285257 & 17$^h$46$^m$21$^s$.578 & -28$^\circ$52\arcmin56\arcsec.40 & 1.36$_{-0.50}^{+0.58}$ &  6.6$_{-1.7}^{+2.4}$	&
 $0.79_{-0.33}^{+0.13}$		& $1.13$	\\
 3 &CXO\,J174622.7-285218 & 17$^h$46$^m$22$^s$.724 & -28$^\circ$52\arcmin18\arcsec.09 & 0.50$_{-0.30}^{+0.34}$ &  7.8$_{-1.5}^{+1.9}$	&
 $4.73_{-1.75}^{+0.70}$	    & $6.38$	\\
 4 &CXO\,J174623.5-285632 & 17$^h$46$^m$23$^s$.599 & -28$^\circ$56\arcmin32\arcsec.08 & 0.37$_{-0.69}^{+0.84}$ & 14.3$_{-5.4}^{+7.5}$	&
 $2.06_{-0.84}^{+1.43}$		& $3.00$	\\
 5 &CXO\,J174611.3-285346 & 17$^h$46$^m$11$^s$.218 & -28$^\circ$53\arcmin47\arcsec.16 & 0.96$_{-0.59}^{+0.66}$ & 16.0$_{-4.7}^{+5.6}$	&
 $1.89_{-0.92}^{+0.25}$		& $3.05$	\\
 & Cloud                 &                        &                                  & 2.12$_{-0.26}^{+0.27}$ &  6.4$_{-0.7}^{+0.8}$	&          	
 $3.48_{-1.46}^{+0.56}$ 	& $5.66$	\\
\hline
\hline
\end{tabular}
\end{minipage}
}
\end{table*}

\begin{figure*}[]
\includegraphics[width=\columnwidth,bb=25 273 568 692,clip]{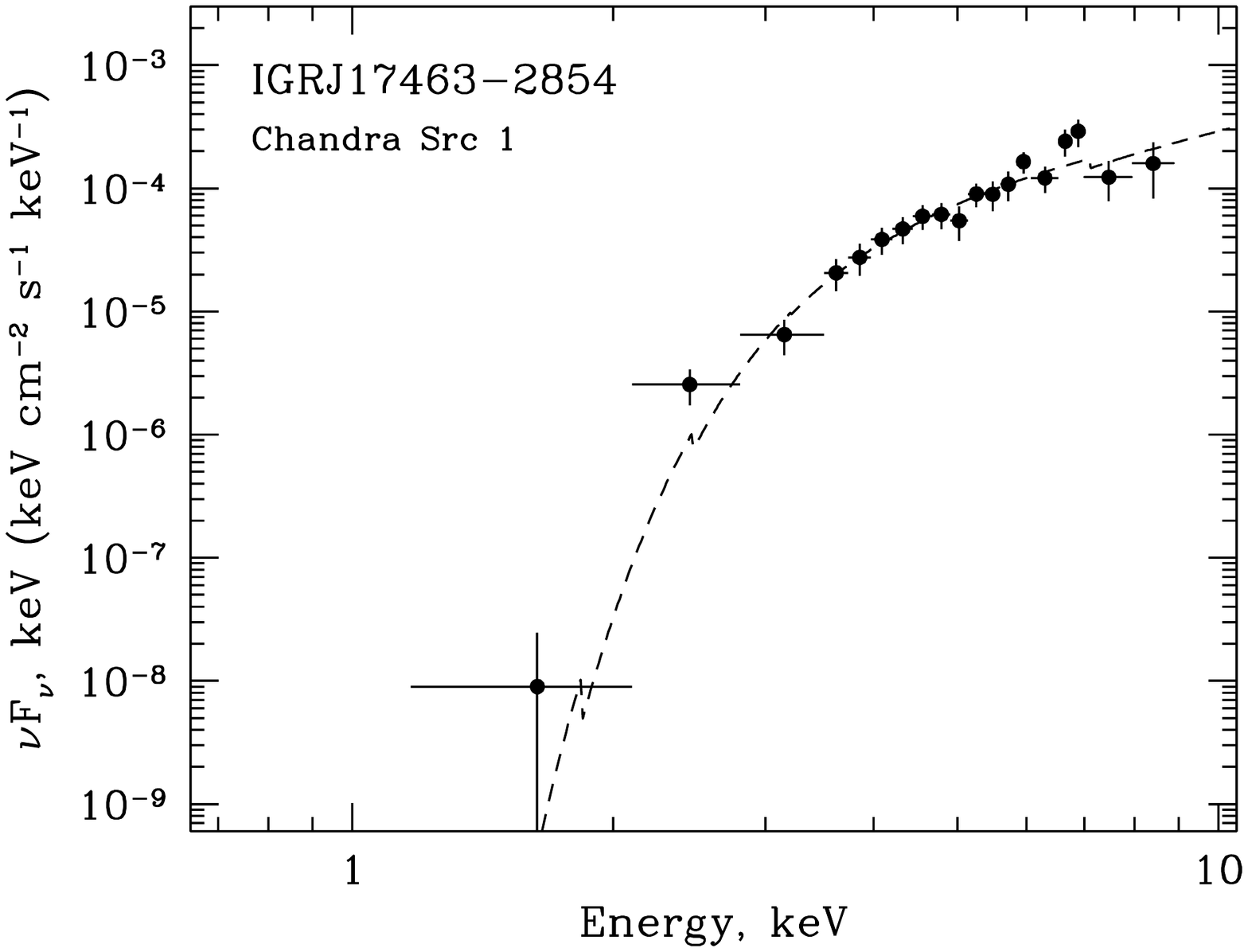}
\includegraphics[width=\columnwidth,bb=25 273 568 692,clip]{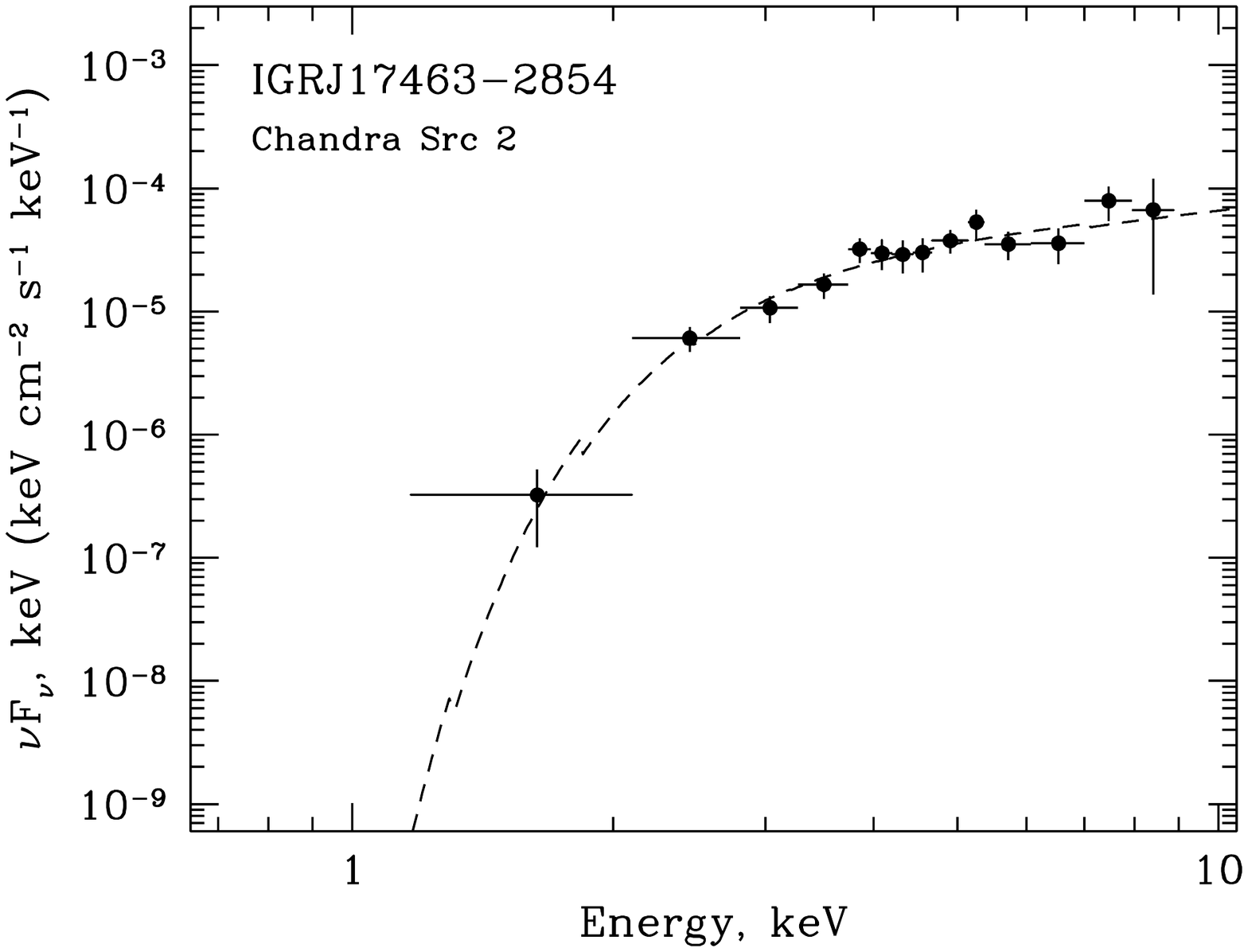}
\includegraphics[width=\columnwidth,bb=25 273 568 692,clip]{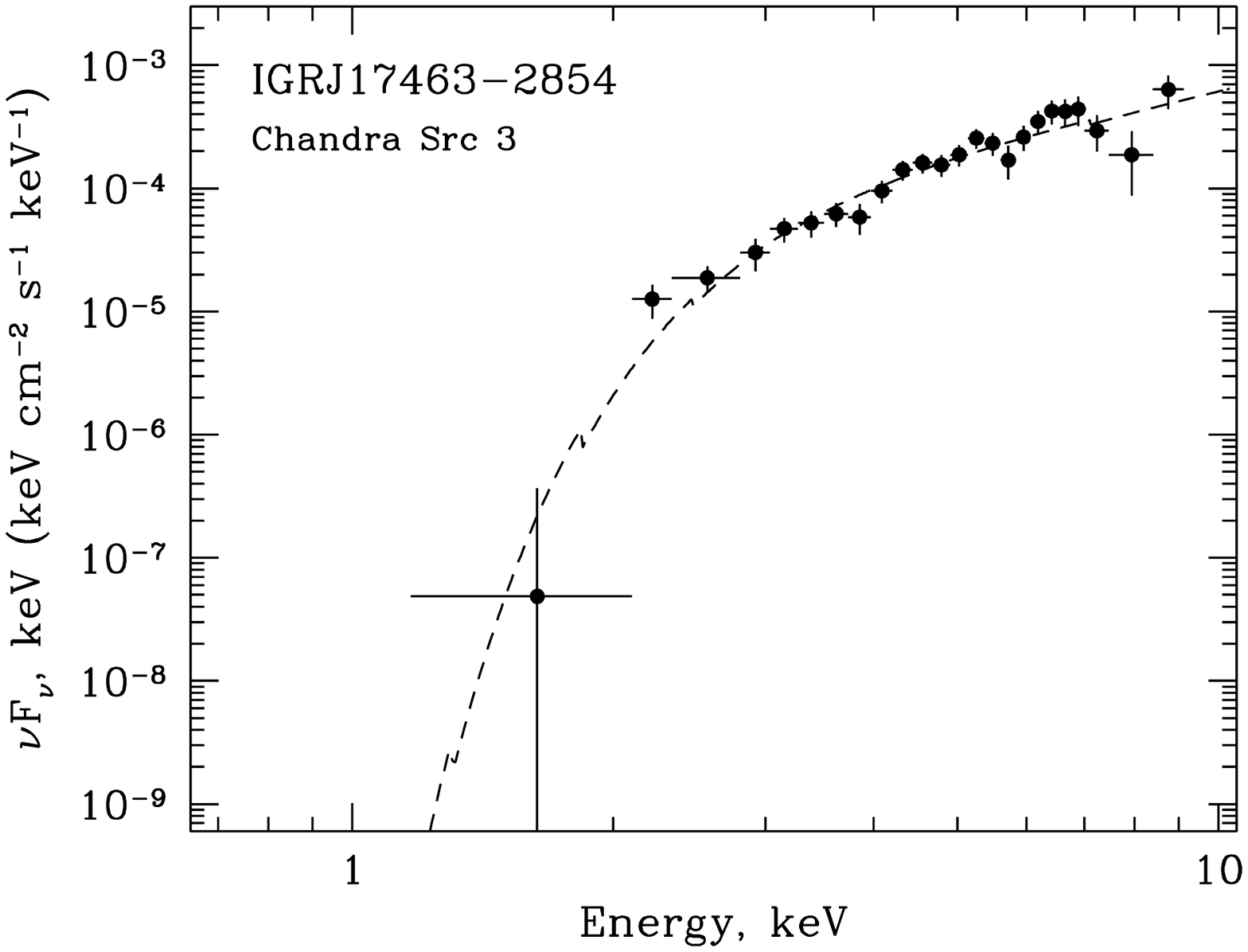}
\includegraphics[width=\columnwidth,bb=25 273 568 692,clip]{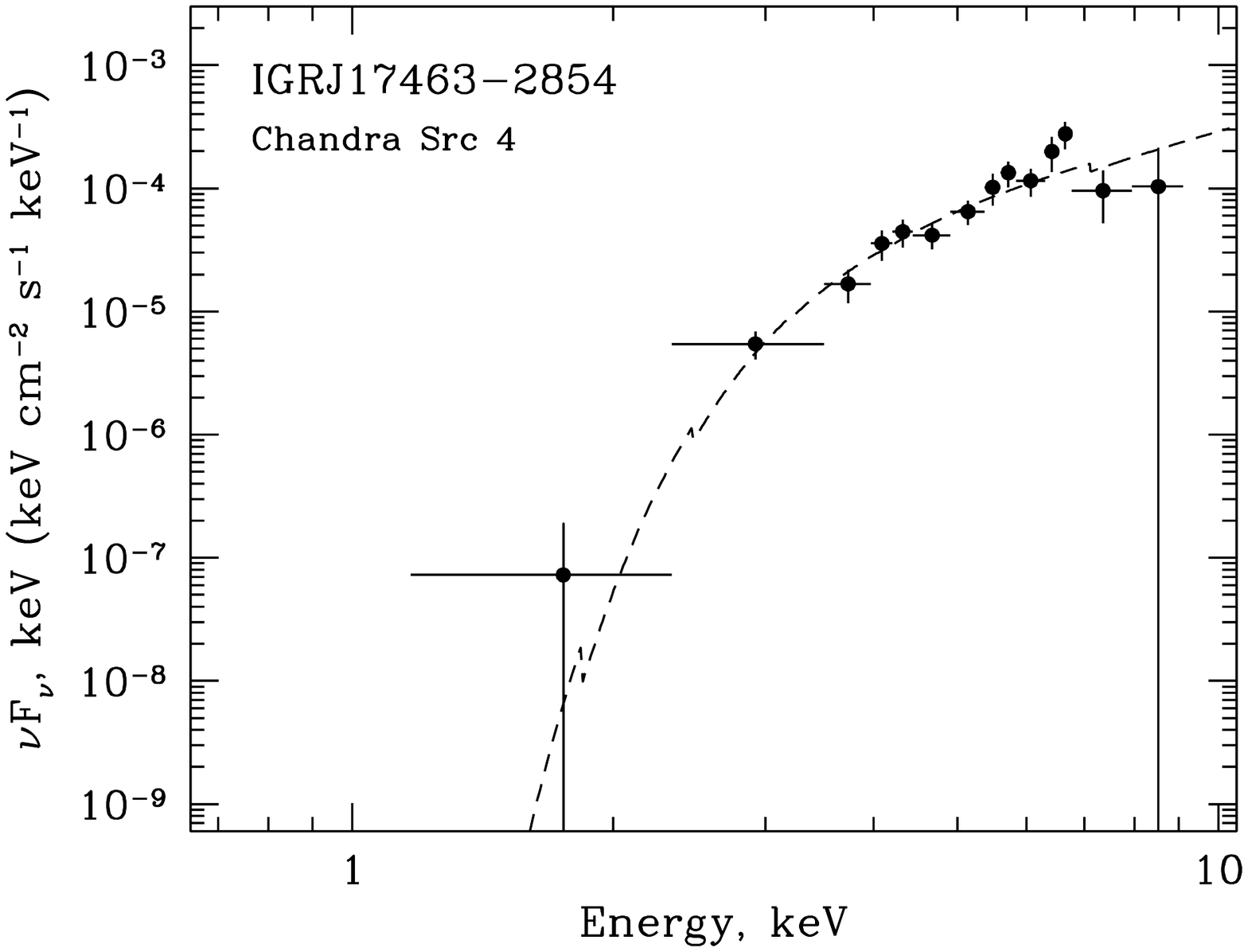}
\includegraphics[width=\columnwidth,bb=25 273 568 692,clip]{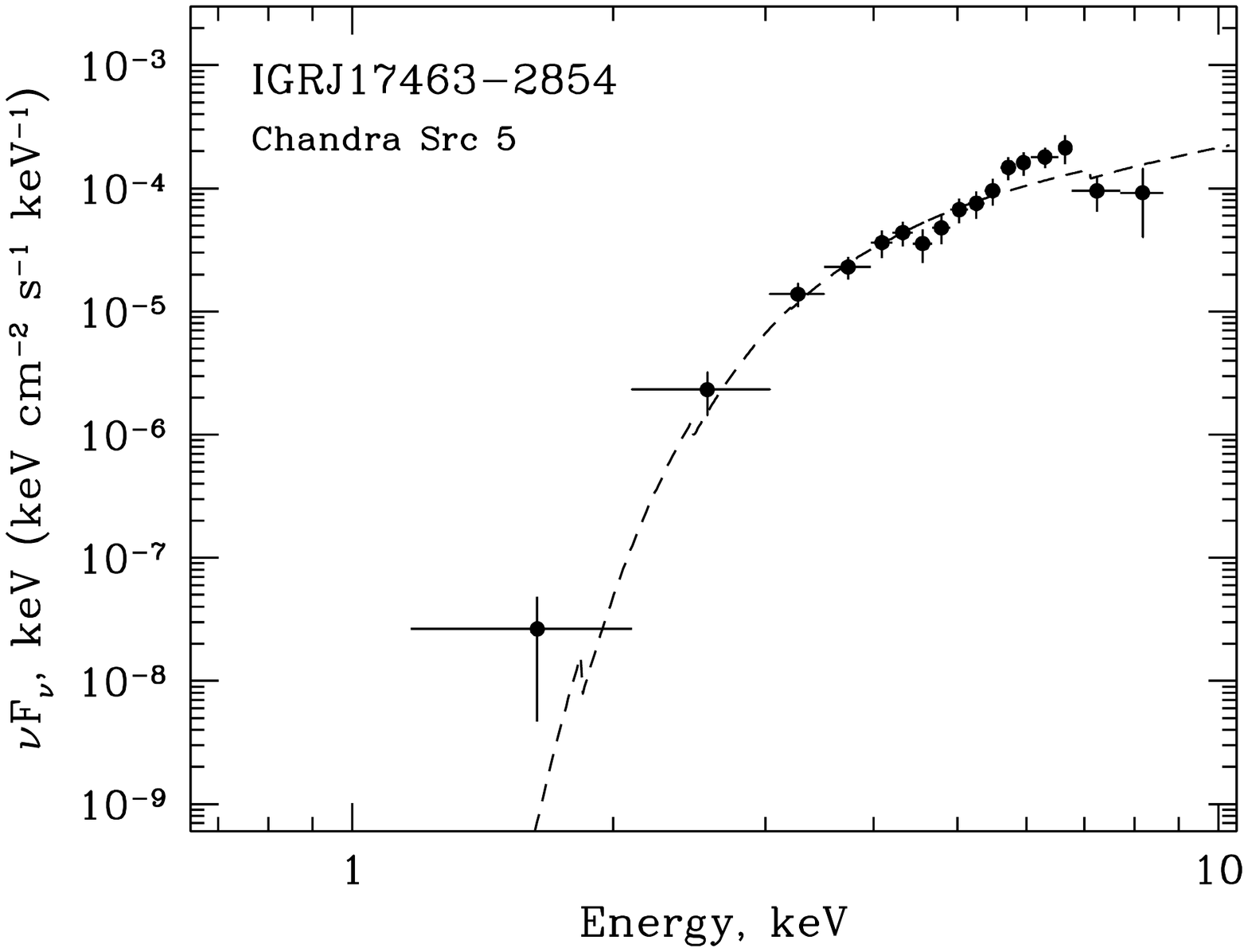}
\hspace{4mm}\includegraphics[width=\columnwidth,bb=25 273 568 692,clip]{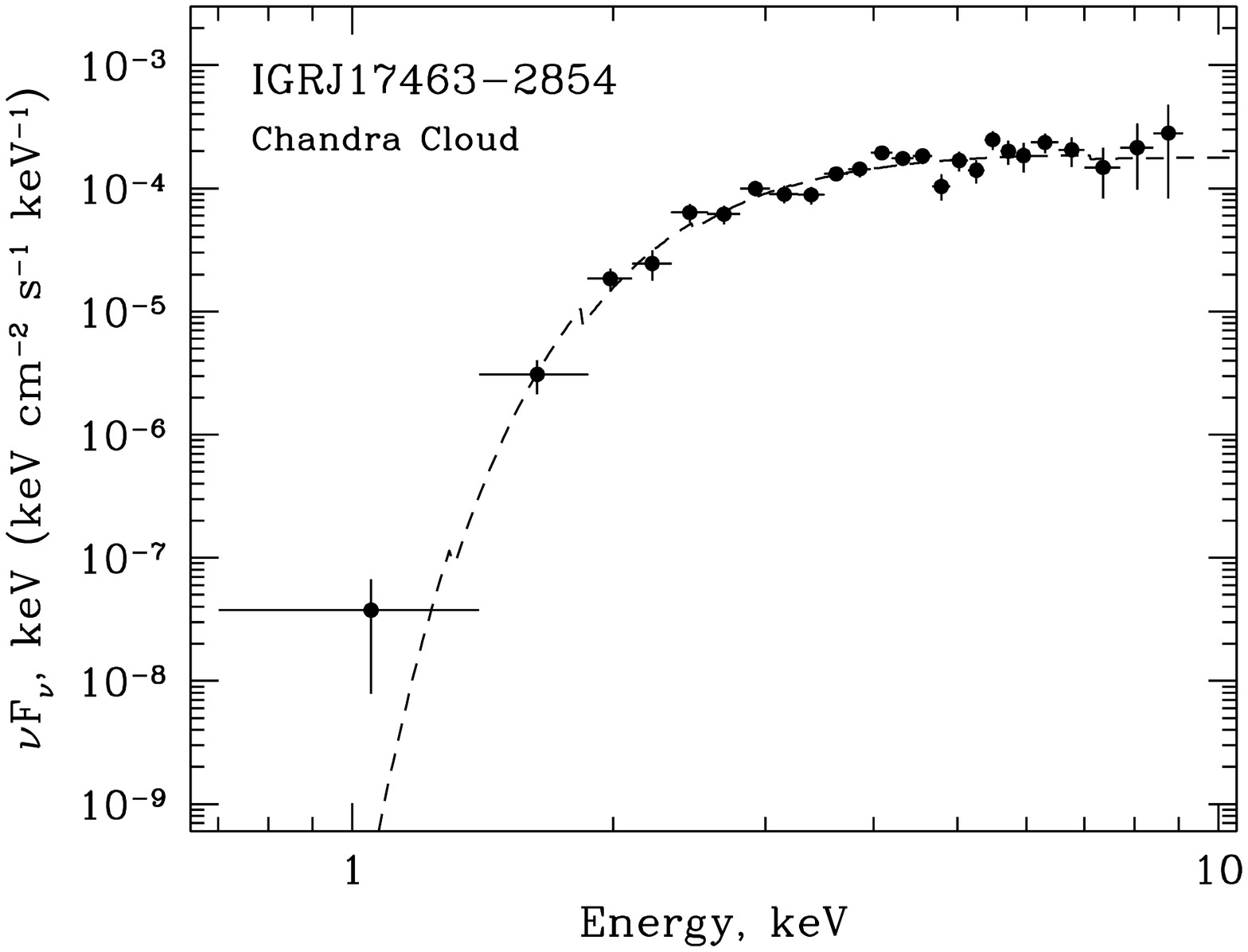}
\caption{\footnotesize{X-ray spectra of the objects detected by Chandra in the error circle of IGR\,J17463-2854. }}  \label{ChSP}
\end{figure*}

However, this approach to estimating the line-of-sight absorption also has shortcomings: there exists
a problem that the application of the standard extinction law is limited for the entire sky, especially
for the Galactic bulge. In particular, it has been shown
in a number of papers that the properties of dust in the Galactic center region can differ from those of
local dust; as a result, applying the standard law can overestimate the absorption by up to a factor
of 2 (Nishiyama et al. 2009; Revnivtsev et al. 2010; Karasev et al. 2010b). Nevertheless, we may consider
the above value as an upper limit on the hydrogen column density for this sky field, and this leads us
to conclude that some of the sources have intrinsic absorption.
As has been said above, the fluxes from all sources are approximately of the same order of magnitude;
nevertheless, source 3 whose flux exceeds the remaining ones approximately by a factor of 2--3 can be
distinguished among them (see Table 1).

\begin{figure*}
\centering
\includegraphics[width=0.95\columnwidth,trim={0.5cm 0 0.2cm 0},clip]{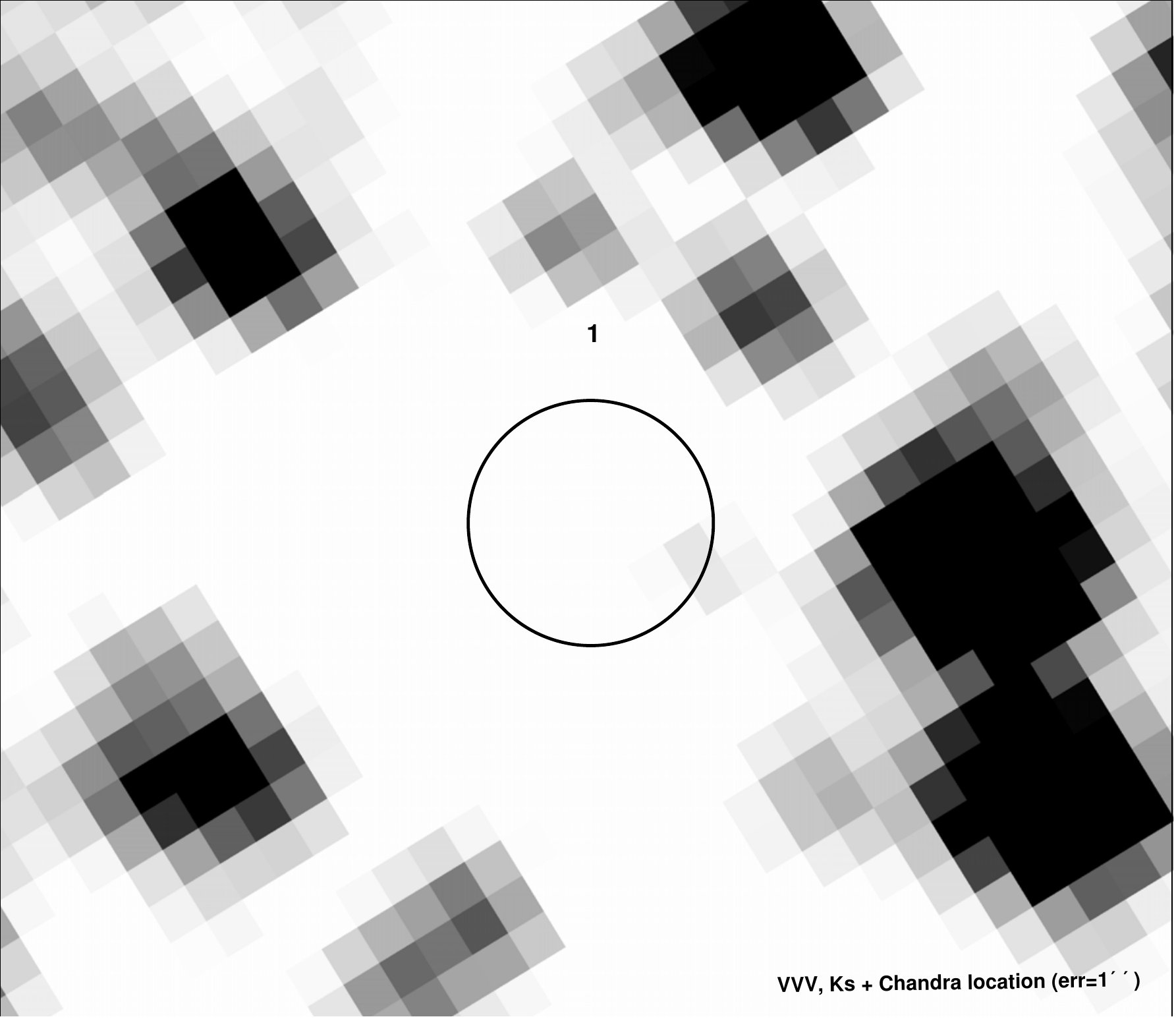}
\includegraphics[width=0.95\columnwidth,trim={0.6cm 0 0 0},clip]{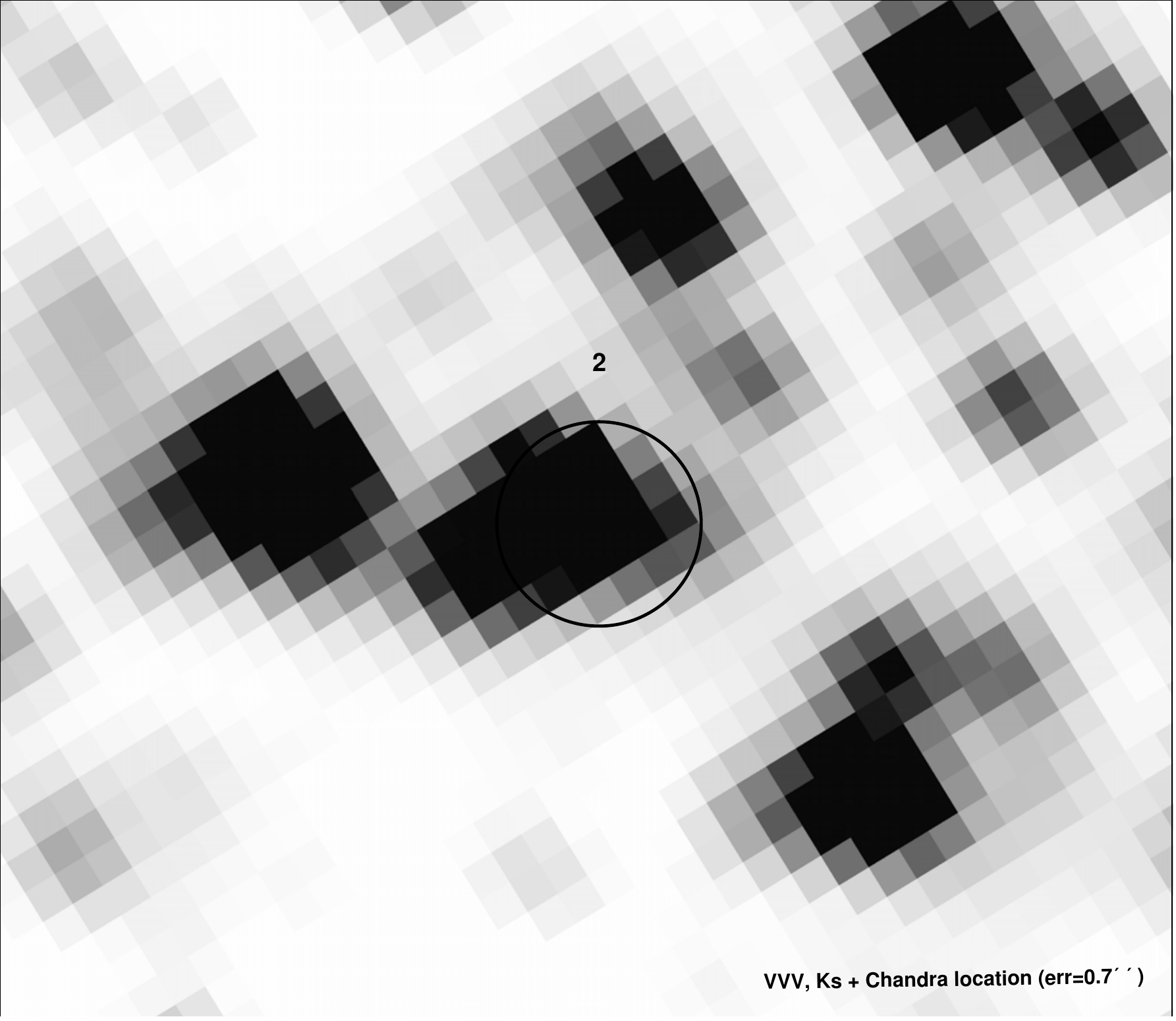}
\includegraphics[width=0.95\columnwidth,trim={0 0 0 0},clip]{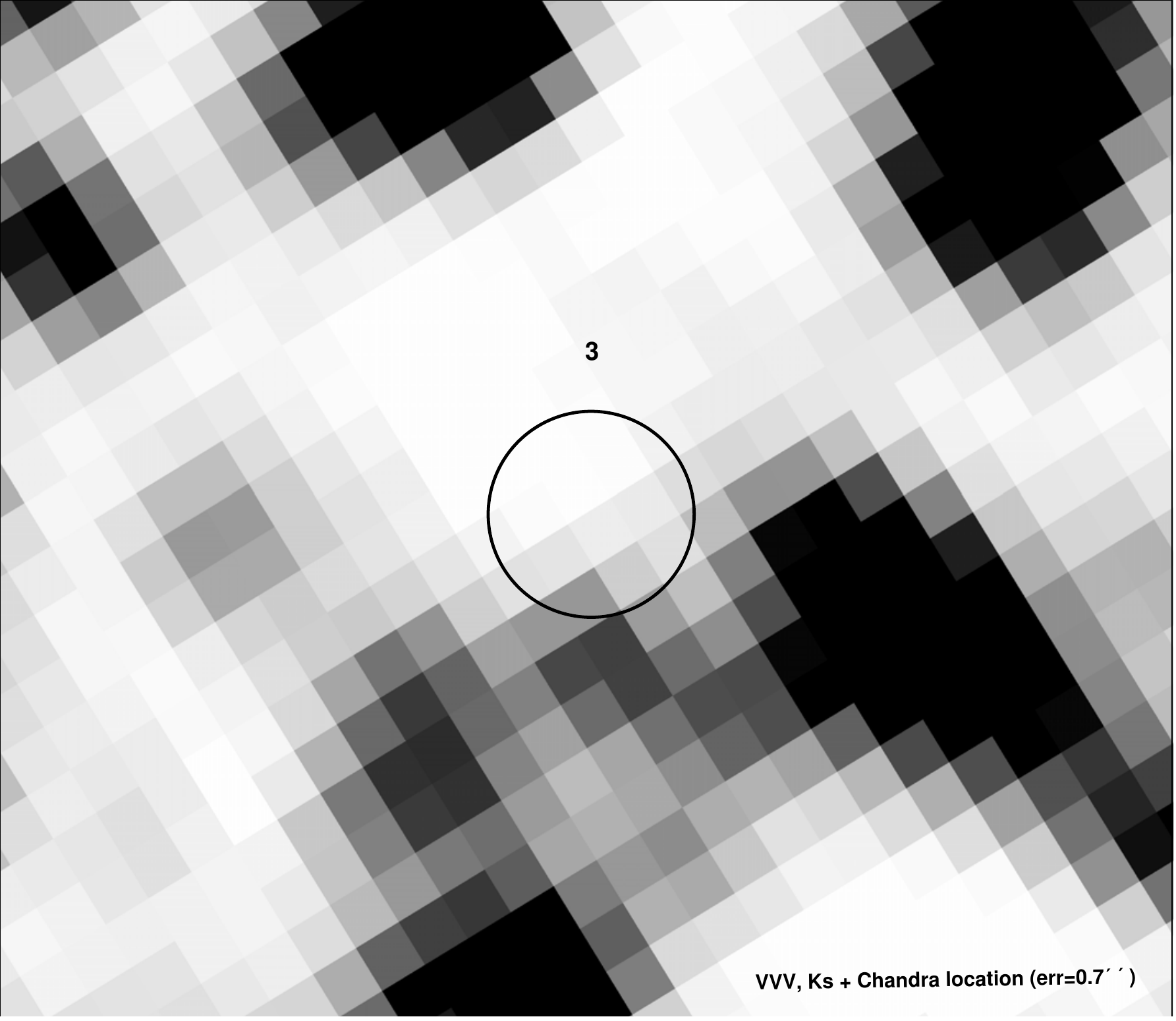}
\includegraphics[width=0.95\columnwidth,trim={0 0 0 0},clip]{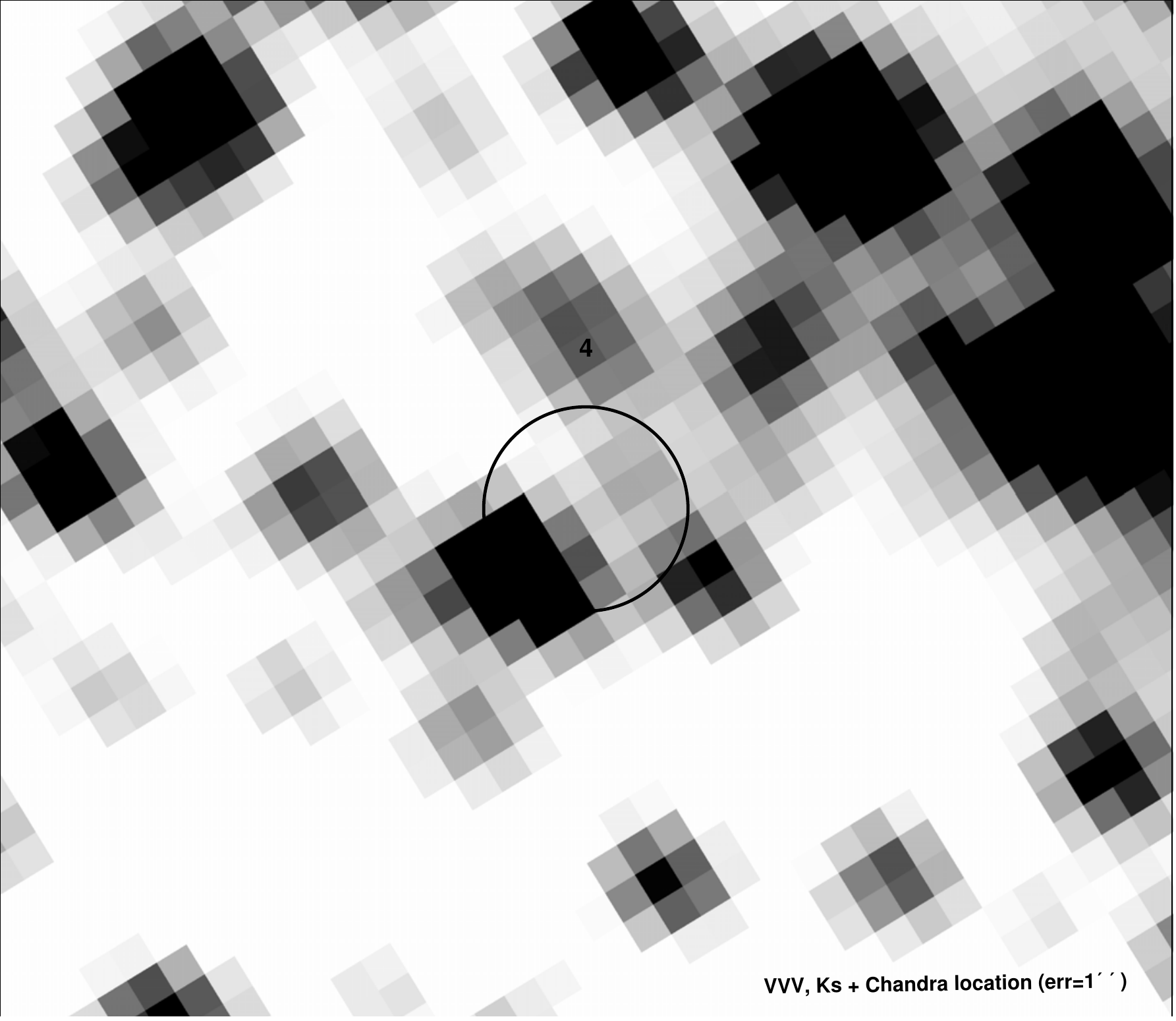}
\includegraphics[width=0.95\columnwidth,trim={0 0 0 0},clip]{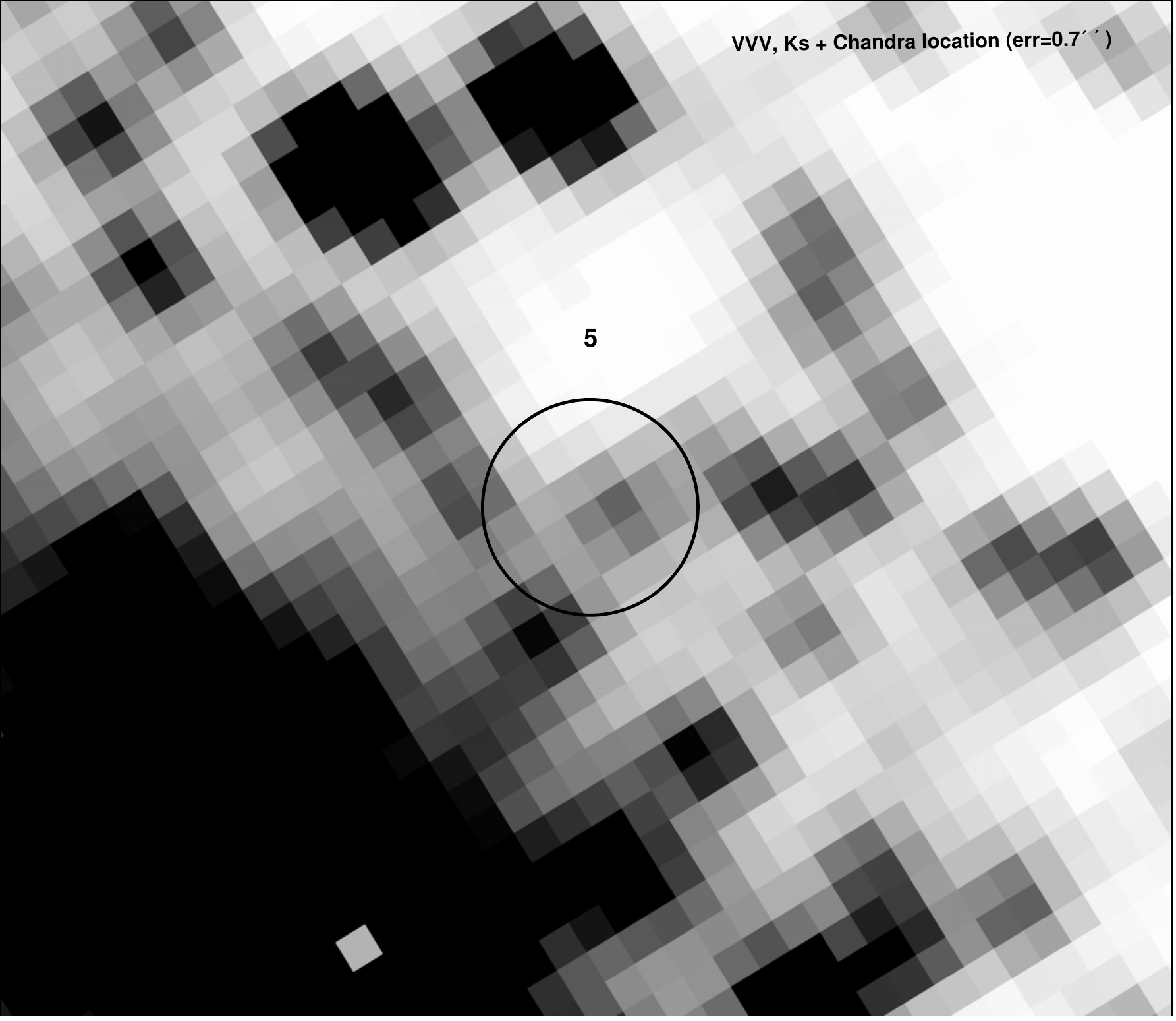}
\caption{\footnotesize{Image of the sky around the X-ray objects detected by Chandra in
the error circle of IGR\,J17463-2854 in the infrared $Ks$ band based on data from the {\it VVV}
survey. The circles correspond to the Chandra localization accuracy of these objects
($0.7-1$\arcsec, specified in the images).}}  \label{VVV}
\end{figure*}

The extended object, the so-called ''cloud'' located
near source 2 and even partially covering it, should be noted separately. The total flux from this
cloud in the 2--10 keV energy band is comparable to the flux from source 3 and, together with source 2,
exceeds it. Unfortunately, the available data do not allow us to answer the question of whether this cloud
and source 2 are associated or are independent objects. Note that the absorption values obtained by
fitting the spectra for these objects are virtually equal. Deviations of the measured fluxes from the model
spectra are observed in the spectra of sources 1, 4, and 5 in the 6--7 keV energy band (Fig. 4), which may be
related to the presence of iron lines at energies 6.4--7.0 keV in the spectra of these objects. The latter may
suggest that they are compact objects, in particular, white dwarfs. Including the emission line at these
energies in the model improves the spectrum fitting quality only slightly, with the upper limits ($1\sigma$)
on the equivalent width of such a line being rather significant, from -- 0.5 keV for source 5 to 0.6 and 0.84
keV for sources 1 and 4, respectively. Note that such iron line equivalent widths can be observed from
white dwarfs in binary systems, for example, from symbiotic stars (see, e.g., Eze 2013).


\section*{THE INFRARED BAND. $VVV$ SURVEY RESULTS}

To determine the nature of the soft X-ray sources detected by Chandra and, accordingly, the possible
nature of IGR\,J17463-2854, we identified them in the infrared energy band based on data from the {\it VVV}
survey. It can be clearly seen from Fig. 5 that only sources 2, 4, and 5 have a statistically significant
identification (their infrared magnitudes are given in Table 2), while sources 1 and 3 do not have such an
identification. Since our attempts to identify them using the UKIDSS catalog did not yield any results
either, only upper limits on the magnitudes are given for them.

\begin{table}
\centering
\footnotesize{
\caption{\footnotesize Infrared magnitudes of the optical counterparts to the sources detected by Chandra}
\begin{tabular}{lccc}
\hline
№	    & $m_J$  			& 	$m_H$			& $m_{Ks}$			\\
\hline
 1      &  $>20.5 $		    &	  $>19.3$       & $>17.5 $	        \\
 2      & $17.05\pm0.08$	& $13.54\pm0.04$ 	& $11.71\pm0.02$	\\
 3     	& $>20 $			&		$>18.7$		& $>17$		        \\
 4     	& $13.59\pm0.01$	&  $13.03\pm0.01$ & $12.80\pm0.02$	\\
 5     	& $>19.3 $			&  $17.06\pm0.11$ & $15.03\pm0.06$	\\
\hline
\end{tabular}
}
\end{table}

Subsequently, we estimated the classes of the counterparts for sources 2, 4, and 5 based on the approach
and the technique from Karasev et al. (2010a) and by taking into account a number of additions
related to the peculiarity of the case under consideration.
The main idea of this technique is as follows: having the photometric observations of the source
located toward the Galactic bulge at least in two filters and knowing the distance and extinction to it
(to be more precise, to the red clump giants that are believed to concentrate there), we can identify
the classes of the stars that can potentially be counterparts of the X-ray source. It should be noted that
variability of the distance to the localization center of the red clump giants (RCGs) along the Galactic
bulge is pointed out in a number of papers (see, e.g., Nishiyama et al. 2009; Gonzalez et al. 2012). In the
case under consideration, this is not so important, because the objects under study are located toward
the Galactic center and the distance to the bulge RCGs in this direction can be assumed to coincide
with the distance to the Galactic center. According to the estimates from a number of papers (Paczynski
and Stanek 1998; Popowski 2000; Udalski 2003; Revnivtsev et al. 2010; Karasev et al. 2010b), it is
determined in the range from 8300 to 8700 pc. For the subsequent analysis, we took the distance to the
Galactic center to be $D_{GC}=8400\pm400$ pc.

To determine the extinction to the Galactic center,
we investigate the positions of RCGs on the color--magnitude diagram constructed for all stars in some
neighborhood ($\sim$1\arcmin $\times$ 1\arcmin) of the object under study. A broadening and a
local increase in the star density typical of the RCGs are usually detected on the red giant branch at a
statistically significant level. Our analysis showed that for the entire sky field where the sources of interest
to us are located, the positions of RCGs on the diagram are virtually invariant and are determined by
the following parameters: $m_{Ks, RCG}=14.60\pm0.08$,  $m_{H, RCG}-m_{Ks, RCG}=1.76\pm0.15$.

The constructed color-magnitude diagrams give a primary idea of the possible class of the counterpart
for some sources (see the left panels in Figs. 7, 8, and 9). In particular, the most probable optical counterpart
to source 2 is a fairly bright red giant, because it lies above the clump on the diagram; the infrared
counterpart to source 5 is very faint, and all estimates for it have a low significance. However, it also falls,
within the error limits, into the region of red giants, but already less bright ones. It can be said about
the optical counterpart to source 4 that it is not a red giant.

\begin{figure}
\centering\includegraphics[width=\columnwidth,bb=26 555 289 806]{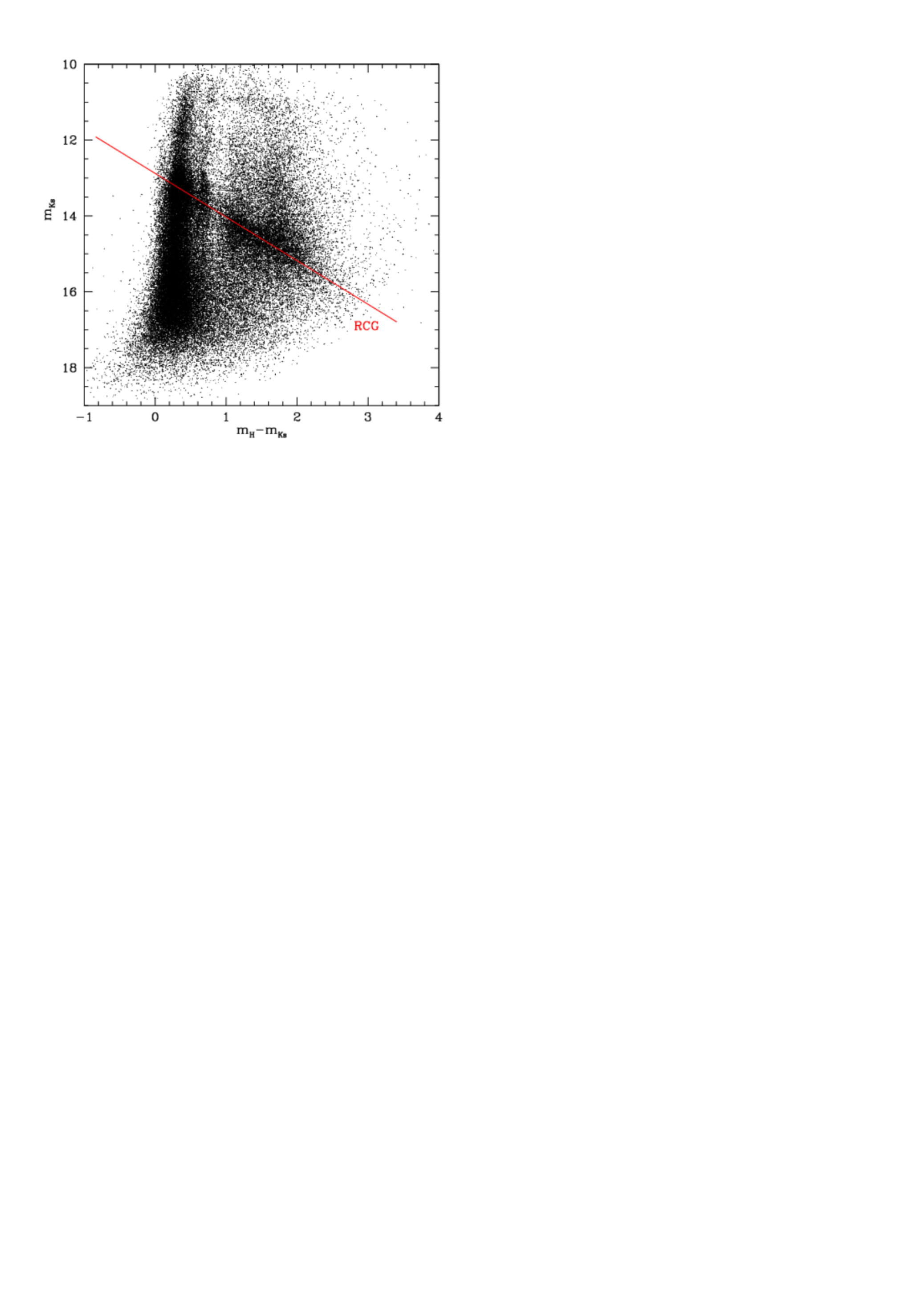}
\caption{\footnotesize{Cumulative color–magnitude diagram constructed for a 2\arcmin\-wide
strip symmetric relative to the Galactic plane (|l|<1 deg.) in the vicinity of source 2. The line
corresponds to the best fit to the RCG branch by a straight line.}\label{STRIPE}}
\end{figure}

Returning to the search for and correct determination of the positions of RCGs on the color–magnitude
diagram, we will note that the clump is fairly close to the detection limit of the {\it VVV}
survey for the sky fields
under consideration. The latter raises the question as to whether these are actually RCGs or these are
just a broadening in the range of faint magnitudes typical for such diagrams. Therefore, we performed
an additional study aimed at improving the positions of RCGs on the color–magnitude diagram in the
region of IGR J17463-2854. Since the RCGs in the direction under consideration are at approximately
equal distances from the observer, their positions on the color–magnitude diagram depend only on the
magnitude and interstellar extinction law (the age and metallicity affect weakly the luminosity of objects
from this class; Paczynski and Stanek 1998).

The latter allows us to reveal a local dependence
of the observed RCG magnitude on color by investigating the photometric properties of giants in some
neighborhood of IGR\,J17463-2854 and to improve the positions of RCGs on the diagram for the region
under study based on the derived dependence. A narrow strip (2\arcmin\ in width) perpendicular to the Galactic
plane and covering the range of latitudes from $-1$ to $+1$ deg. is best suited as such a neighborhood. Owing
to the choice of such a region, we can minimize the possible scatter of distances related to the variability
of the bulge structure (Nishiyama et al. 2009; Gonzalez et al. 2012).

Figure 6 presents a cumulative color–magnitude diagram for such a strip including the sky field under
study. The region of an enhanced star density formed by RCGs is clearly seen on this diagram.
The positions of giants can be improved by fitting the derived RCG branch. In addition, the slope of
this straight line allows us to determine the extinction law $A_{Ks}/E(H-Ks) = 1.12\pm0.11$ toward the
bulge for the sky field under study (see also Karasev et al. 2010b), which differs significantly from its
standard value of $\approx1.8$ (Cardelli et al. 1989; Schlegel et al. 1998).

Our investigation confirms the validity and correctness of our previous magnitude and color estimates
for RCGs toward IGR J17463-2854. Thus, the Ks extinction to the Galactic center in this
direction is $A_{Ks,GC}=1.60\pm0.17$, where $A_{Ks,GC}=m_{Ks,RCG}-M_{Ks,RCG}+5-5{\rm log} (D_{GC})$,
$M_{Ks,RCG}=-1.61 \pm 0.05$  and $M_{Ks,RCG}=-1.61 \pm 0.05$ is the absolute magnitude of RCGs (Alves 2000;
Karasev et al. 2010a).

Using this information, we can estimate the class of the counterparts to the sources under study as
follows. Considering stars of different spectral types and luminosity classes as a counterpart (by simple
exhaustion), we can determine what correction for the extinction ($A_{Ks,{\rm test}}$) and distance
($D_{\rm test}$) is required for a star of each of the classes to satisfy the actual observed magnitudes
from Table 2.

$A_{Ks,{\rm test}}$ is determined by comparing the unabsorbed color of a test star $(M_H-M_{Ks})_{\rm test}$
with the color of a real star ($(m_H-m_{Ks})_{\rm real}$) for a known extinction law ($R$), which allows the
following quantities in two filters ($A_{H,{\rm test}}$ and $A_{Ks,{\rm test}}$) to be related:

\begin{equation}
A_{H,{\rm test}}-A_{Ks,{\rm test}} = (m_H-m_{Ks})_{\rm real}-(M_H-M_{Ks})_{\rm test},
\end{equation}
\begin{equation}
A_{Ks,{\rm test}}/(A_{H,{\rm test}}-A_{Ks,{\rm test}}) = R .
\end{equation}

\noindent Given the extinction correction, we determine the necessary distance correction:
\begin{equation}
5log(D_{\rm test}) = m_{Ks,{\rm real}}-M_{Ks,{\rm test}} +5 - A_{Ks,{\rm test}}.
\end{equation}

In the next step, we check how these corrections correspond to the previously determined extinction
and distance to the Galactic center, our unique reference. If the necessary extinction correction for a
star of some class is larger than the extinction to the Galactic center ($A_{Ks,{\rm test}} > A_{Ks,GC}$),
while the distance correction is smaller than the distance to it ($D_{\rm test} < D_{GC}$), then such a
star cannot be a counterpart to the source under study in view of a clear mutual contradiction between
the corrections. The reverse is also true: if $A_{Ks,{\rm test}} < A_{Ks,GC}$ and $D_{{\rm test}} > D_{GC}$,
then such a star cannot be a counterpart either. The absolute magnitudes of stars of various classes
$M_{Ks,{\rm test}}$ were taken from Wegner (2007, 2014).

It is necessary to note two important (key) points. First, since we make estimates simultaneously for
the extinction and distance, we need to know the observed photometric magnitudes of a star at least in
two filters (desirably weakly affected by metallicity) and the extinction law. As has been shown above,
the extinction law for RCGs in the direction under consideration differs significantly from the standard
one. However, the standard extinction law is valid at close distances (up to several kpc) (Marshall
et al. 2006). Since we do not know where and precisely how the extinction law changes, we used
the following simplifying assumption: the standard law $R = A_{Ks}/E(H-Ks) = 1.8$ is used closer to
the Galactic center ($D < 8400$ pc) and the cumulative nonstandard law $R = A_{Ks}/E(H-Ks) = 1.12$
is used for sources at and behind the Galactic center.

\begin{figure*}
\centering
\includegraphics[width=0.95\columnwidth,trim={0cm 0.0cm 0cm 0cm},clip]{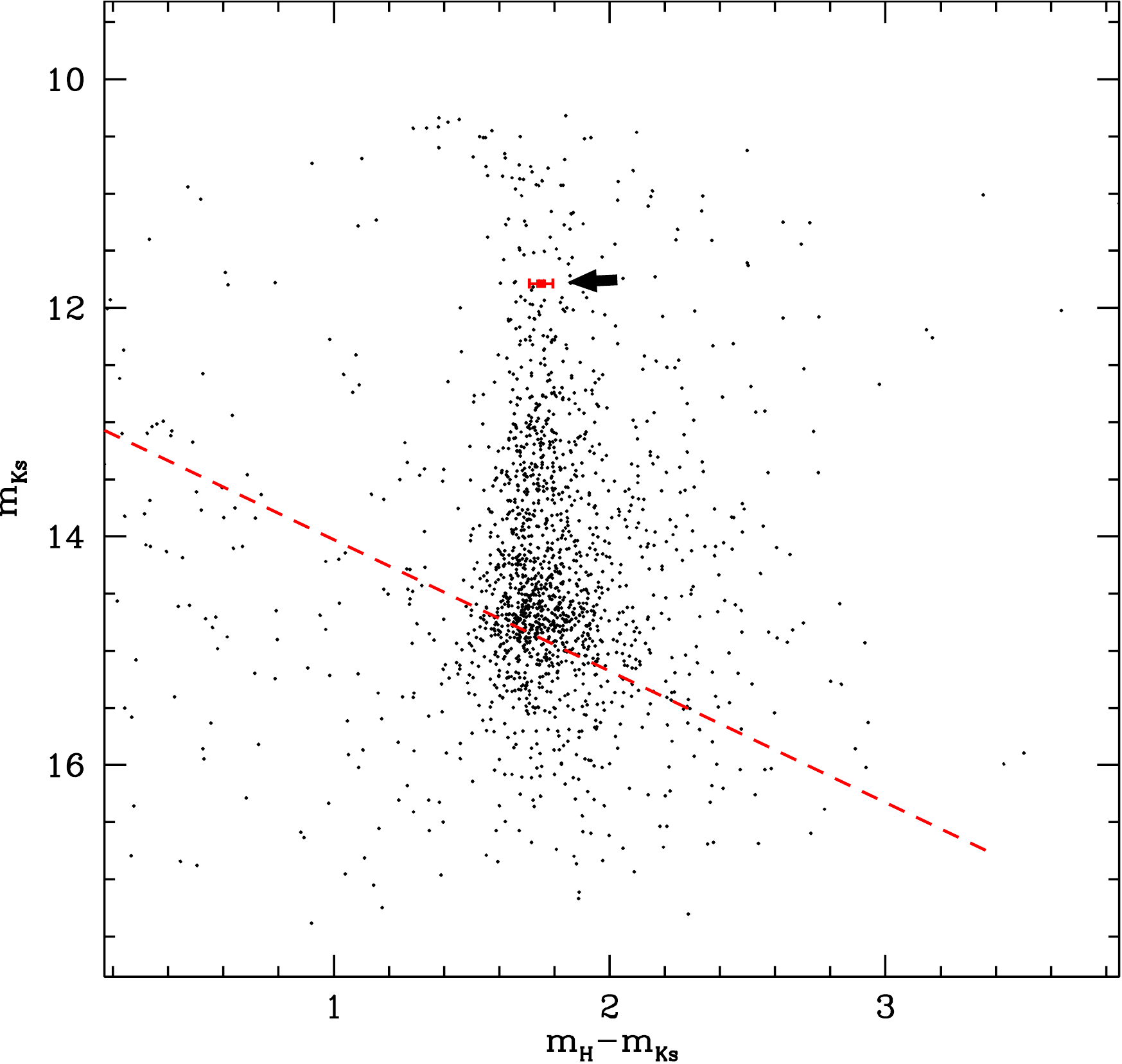}
\hspace{3mm}\includegraphics[width=0.95\columnwidth,trim={0cm 0cm 0 0cm},clip]{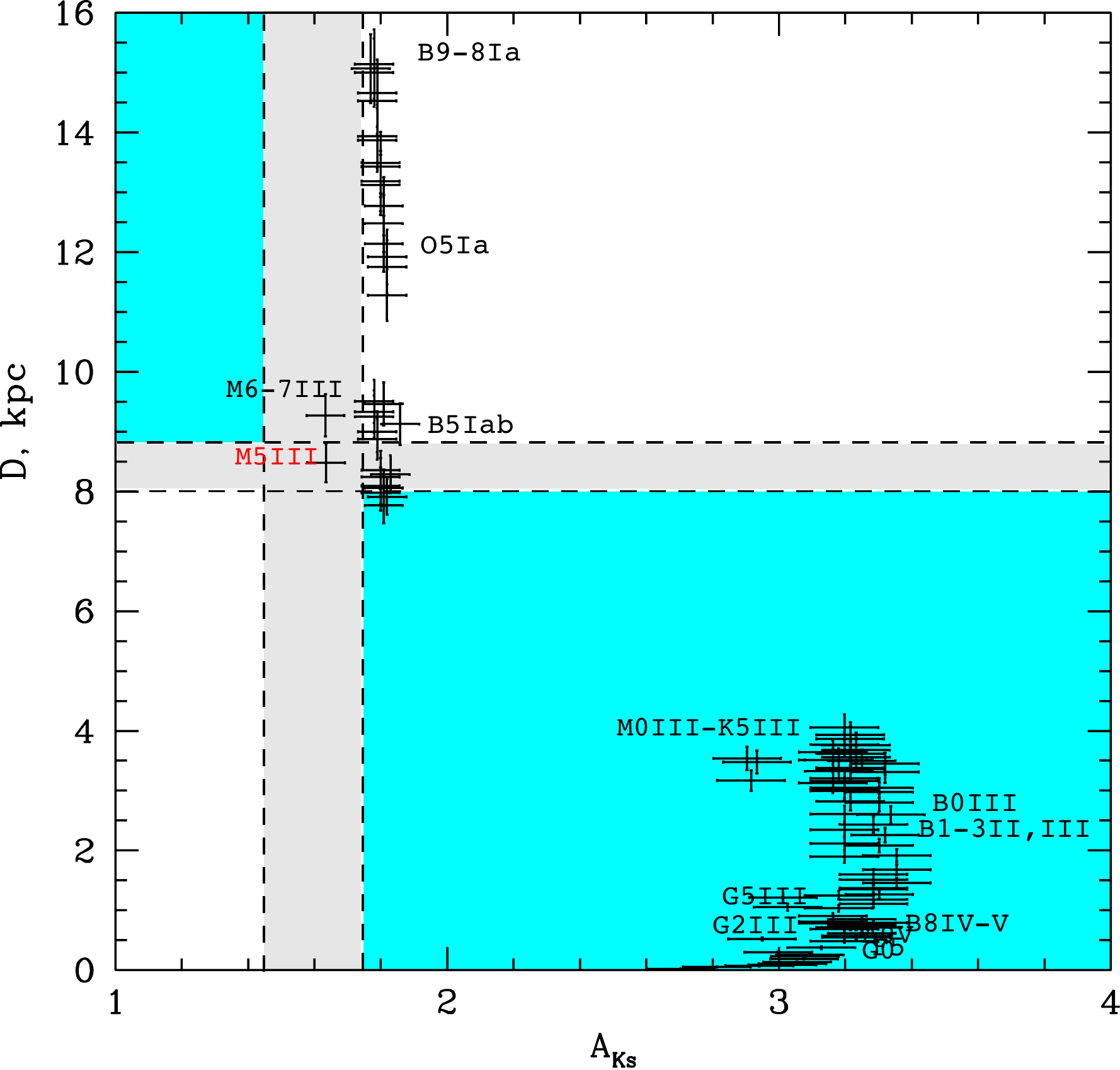}
\caption{\footnotesize{Left panel: the color–magnitude diagram constructed for a 1\arcmin\
neighborhood of source 2; its position is indicated by the square and arrow. The dashed line
corresponds to the solid line in Fig. 6. Right panel: the extinction correction–distance
correction diagram showing the distance at which the counterpart to the source of the
corresponding class would be located and what extinction it would have. The permitted and
''forbidden'' classes lie in the white and blue (gray) regions of the diagram,
respectively. The light-blue (light-gray) region and the dashed lines reflect the inaccuracy
of our knowledge of the distance and extinction to the Galactic center.}}\label{DAK2}
\end{figure*}

The second key point is how to understand for which classes of stars we should choose the standard law and for which the nonstandard one, i.e., how to
understand where a star of some class should have been located before the investigation described above.
To answer this question, we used the following simple test. Going sequentially through stars of different
classes, we add the distance and extinction corrections corresponding to the Galactic center to their
absolute magnitudes and determine what ''apparent'' $Ks$ magnitude would be obtained for each tested star,
i.e.,$m_{Ks,{\rm test},GC} = M_{Ks,{\rm test}} + 5{\rm log}(D_{GC})-5+A_{Ks,GC}$. Then, we compare this magnitude with the
actually observed $Ks$  of the counterpart of the source ($Ks$ ($m_{Ks,{\rm real}}$ from Table 2). If $m_{Ks,{\rm test},GC} <
m_{Ks,{\rm real}}$, then the corresponding corrections for the Galactic center are not enough to sufficiently attenuate
the star of the chosen class. It turns out that such a star could be a counterpart to the source only
if it were farther from the Galactic center. The reverse is also true if $m_{Ks,{\rm test},GC} < m_{Ks,{\rm real}}$,real. Note that the
photometric data only in one filter are sufficient for such qualitative estimates of the distance to the star;
for the same reason, a change in the extinction law does not affect these estimates.

\begin{figure*}
\centering
\includegraphics[width=0.95\columnwidth,trim={0cm 0.0cm 0cm 0cm},clip]{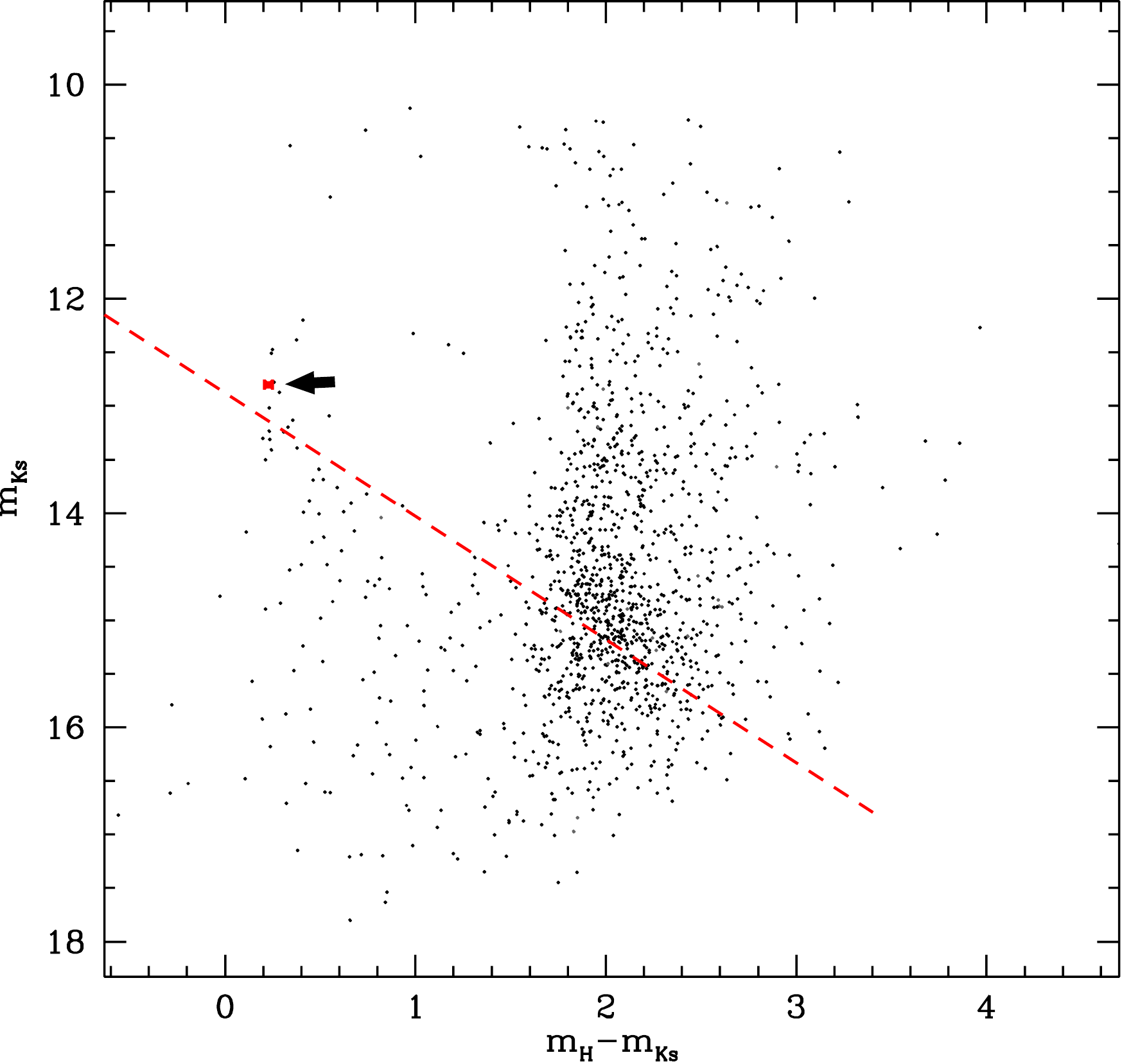}
\hspace{3mm}\includegraphics[width=0.95\columnwidth,trim={0cm 0cm 0 0cm},clip]{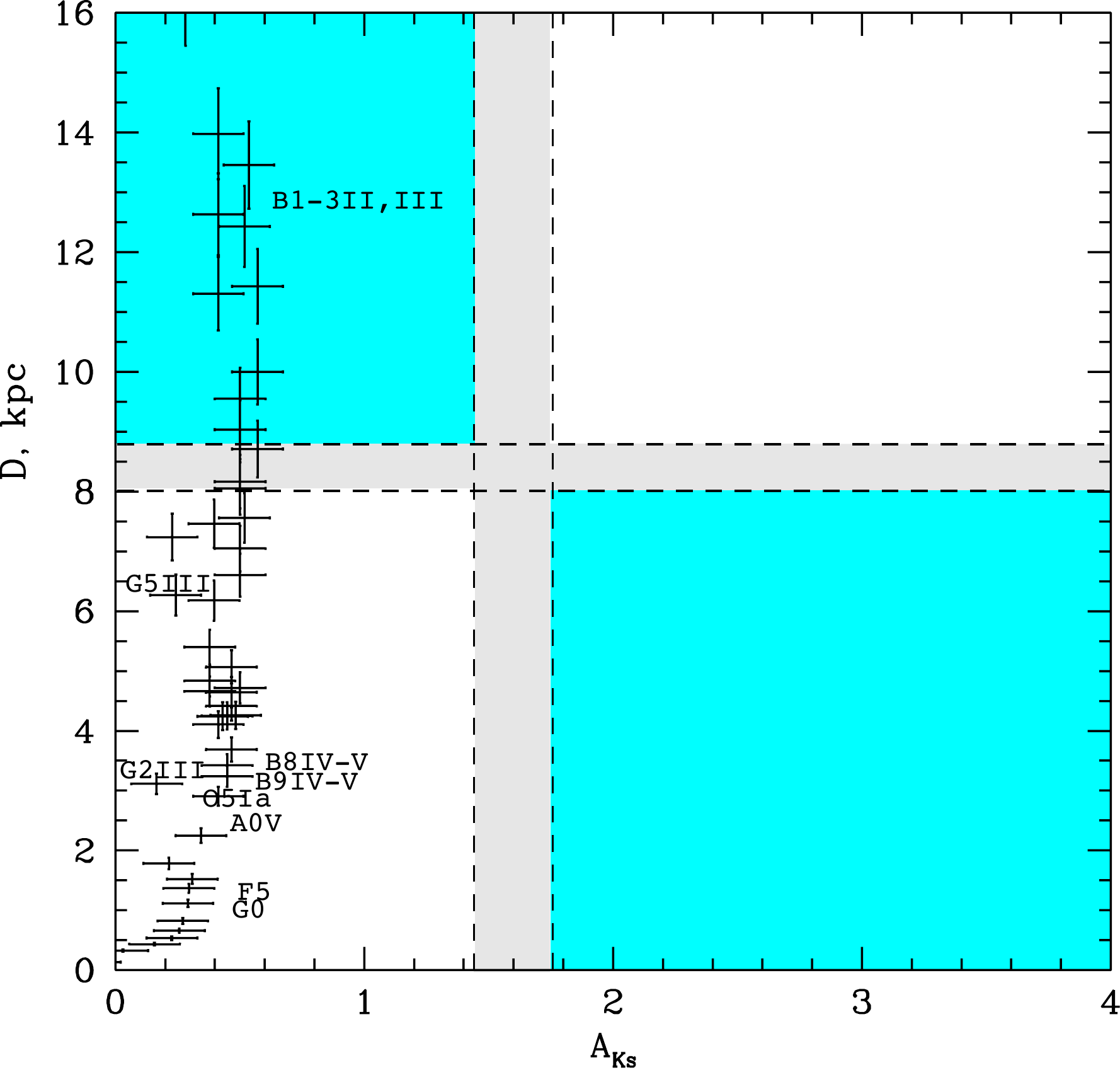}
\caption{\footnotesize{Same as Fig. 7 for source 4.}}\label{DAK4}
\end{figure*}

\begin{figure*}
\centering
\includegraphics[width=0.97\columnwidth,trim={0cm 0.0cm 0cm 0cm},clip]{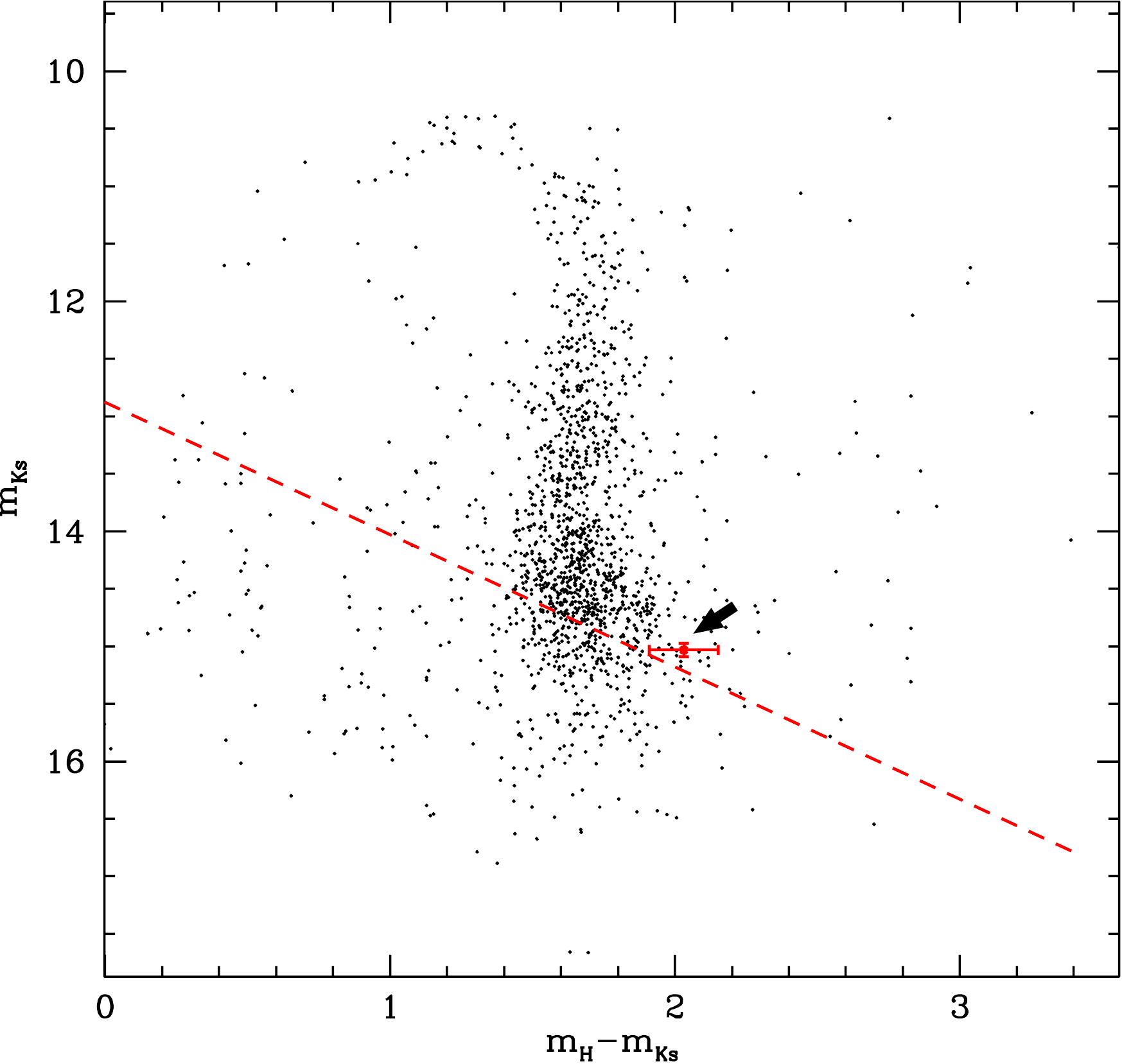}
\includegraphics[width=1.03\columnwidth,trim={0.3cm 1.8cm 0 0cm},clip]{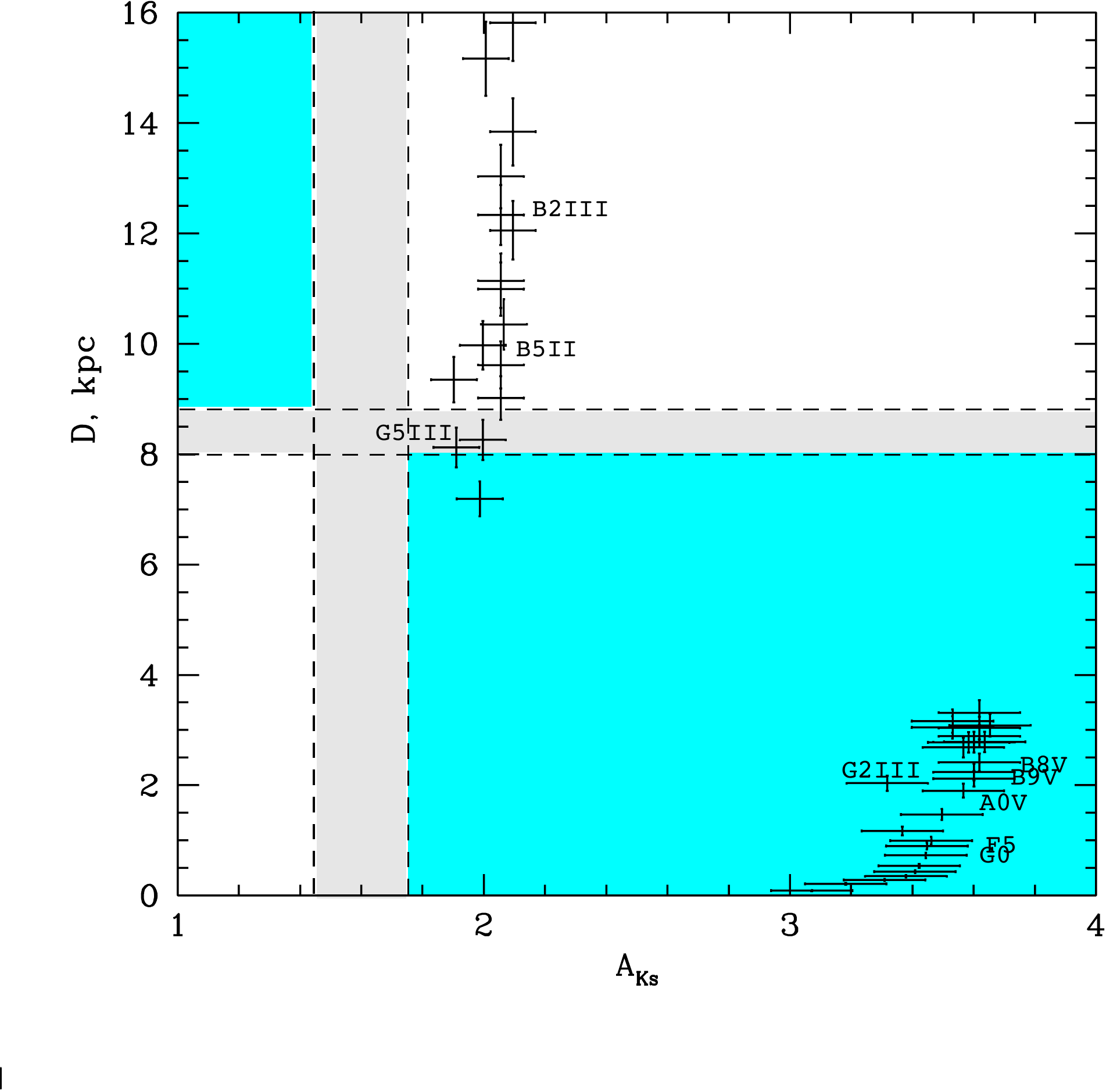}
\caption{\footnotesize{Same as Fig. 7 for source 5.}}\label{DAK5}
\end{figure*}

Having determined the boundaries of applicability of a particular extinction law and applying the
technique described above, we can estimate what classes of stars the optical counterparts to sources 2,
4, and 5 most likely are. The results are presented in the form of distance correction--extinction correction
diagrams on which the regions of forbidden and permitted combinations are marked by the blue (gray)
and white colors, respectively (Figs. 7, 8, and 9, right panels). Note that the ''jump'' of points
on the diagrams is explained by the application of different extinction laws for stars of different classes.

It follows from these diagrams that the optical star in the system with source 2 is most likely an M5III
red giant, although brighter M6-7III stars can also be counterparts to the object because of the errors in
determining the distance to the Galactic center and the corresponding extinction. The optical counterpart
to source 5 may also belong to the class of red giants (approximately G5III), although stars of other
spectral types brighter than B6 cannot be ruled out either, within the error limits (Fig. 9). In comparison
with source 2, the color–magnitude diagram gives no unambiguous indication that it is a red giant of
the bulge. We cannot say anything certain about the class of the optical counterpart to source 4 from the
presented diagram (Fig. 8). Thus, among the five point X-ray sources detected by Chandra in the error
circle of IGR\,J17463-2854, at least one is a possible candidate for a symbiotic binary system.
Note that the described tests and methods are valid under the assumption of a fairly uniform growth
of the cumulative extinction with distance in the absence of dust cocoons around the sources.

\section*{CONCLUSIONS}

In this paper, we investigated in detail the properties of the poorly studied hard X-ray source
IGR\,J17463-2854. Improving the error circle of this object allowed us to identify it in other energy
bands. In particular, according to the Chandra data, five point and one extended objects of comparable
intensity are detected at a statistically significant level in the error circle of IGR\,J17463-2854.

The total flux registered from all these objects in the
2--10 keV energy band is $\simeq1.5\times10^{-12}$ erg s$^{-1}$ cm$^{-2}$.
If this flux is rescaled to the 25--60 keV energy band under the assumption of broadband spectra typical
of symbiotic stars (Kennea et al. 2009; Eze 2013), neutron stars (Filippova et al. 2005), or active
galactic nuclei projected onto the Galactic plane (see, e.g., Sazonov et al. 2004), then the expected
flux in this band will be considerably lower than that measured by INTEGRAL. Thus, the hard X-ray flux of
$\simeq10$mCrab from IGR\,J17463-2854 cannot be explained by a simple superposition of the persistent
fluxes from the objects detected by Chandra. The observed transient behavior of IGR\,J17463-2854 is most
likely related to the variability of the emission from one of the soft X-ray sources. Their variability
cannot be revealed on time scales of hundreds and thousands of seconds based on the Chandra data, because
the sources are too faint to construct a statistically significant light curve. The observations spaced
almost 10 years apart did not reveal any significant changes in the fluxes either.

Nevertheless, based on the results of our studies, we showed that at least one of the objects detected
by Chandra in the error circle of the hard Xray source IGR\,J17463-2854 is probably a symbiotic
system, which are characterized by noticeable flux variability (Yungelson et al. 1995; Kennea et al.
2009; Eze 2013).
The relativistic object in it is most likely a white dwarf with a luminosity $L_X\simeq8.9\times10^{32}$ \ergs,
while the normal star is an M5--M7III red giant located at the distance of the Galactic
center. We cannot say anything unequivocal about the nature of the remaining soft X-ray sources, but
their X-ray spectra suggest that they can also be white dwarfs.

In conclusion, note that we also tried out the technique of searching for symbiotic systems in the
Galactic bulge based on the results from Karasev et al. (2010a) and improved data on the absorption
and extinction law for the Galactic center region (Karasev et al., in preparation).

\section*{ACKNOWLEDGMENTS}

We thank M.G. Revnivtsev, A.N. Semena, and A.A. Voevodkin for the discussion of our results and
useful remarks. We are also grateful to E.M. Churazov, who developed the IBIS/INTEGRAL data analysis
methods and provided the software. This work was financially supported by the Russian Science Foundation (project no. 14-22-00271).

\section*{REFERENCES}
\noindent
1. D.R. Alves, \apj {\bf 539}, 732 (2000).\\
2. A.J. Bird, A. Bazzano, and L. Bassani, \apjs\ {\bf 186}, 1 (2010).\\
3. R.C. Bohlin, B.D. Savage, and J.F. Drake, \apj\ {\bf 224}, 132 (1978).\\
4. J. Cardelli, G. Clayton, and J. Mathis, \apj\ {\bf 345}, 245 (1989).\\
5. E. Churazov, R. Sunyaev, J. Isern, J.  Knodlseder, P. Jean,  F. Lebrun, N. Chugai, and S. Grebenev, Nature {\bf 512}, 406 (2014).\\
6. I.N. Evans, F.A. Primini, K.J. Glotfelty, C.S. Anderson,N.R. Bonaventura, J.C. Chen, J.E. Davis, and S.M. Doe, \apjs\ {\bf 189}, 37  (2010).\\
7. R. Eze, MNRAS {\bf 437}, 857 (2013).\\
8. E. Filippova, S. Tsygankov, A. Lutovinov, and R. Syunyaev, Astron. Lett. {\bf 31}, 729 (2005).\\
9. O.A. Gonzalez, M. Rejkuba, M. Zoccali, E. Valenti, D. Minniti,    M. Schultheis, R. Tobar, and B. Chen.),  \aap\ {\bf 552}, 9 (2012).\\
10. P. Kalberla, W. Burton, D. Hartmann, E. Arnal, E. Bajaja, R. Morras and W. Poppel, \aap {\bf 440}, 775 (2005).\\
11. D. Karasev, A. Lutovinov, and R. Burenin, MNRAS {\bf 409}, L69 (2010а)\\
12. D.I. Karasev, M.G. Revnivtsev, A.A. Lutovinov, R.A. Burenin, Astron. Lett. {\bf 36}, 788 (2010).\\
13. J. Kennea, K. Mukai, J. Sokoloski, G. Luna, J. Tueller, C. Markwardt and D. Burrows, \apj\ {\bf 701}, 1992 (2009).\\
14. R. Krivonos, M. Revnivtsev, S. Tsygankov, S. Sazonov, A. Vikhlinin, M. Pavlinsky, E. Churazov, and R. Sunyaev, \aap\ {\bf 519}, A107 (2010).\\
15. A. A. Lutovinov, M. G. Revnivtsev, D. I. Karasev,  V. V. Shimanskii, R. A. Burenin, I. F. Bikmaev, V. S. Vorobyov, C. S. Tsygankov, and M. N. Pavlinsky, Astron. Lett. {\bf 41}, 179 (2015).\\
16. D. Marshall, A. Robin, C. Reyle, M. Schultheis, and S. Picaud, \aap\ {\bf 453}, 635 (2006).\\
17. S. Nishiyama, M. Tamura, H. Hatano, D. Kato, T. Tanabe, K. Sugitani, and T. Nagata), \apj\ {\bf 696}, 1407 (2009).\\
18. B. Paczynski and K. Stanek, \aap\ {\bf 494}, 219 (1998).\\
19. P. Popowski, \apj\ {\bf 528}, 9 (2000).\\
20. M. Revnivtsev, M. van den Berg, R. Burenin, J. Grindlay, D. Karasev, and W. Forman, \aap {\bf 515} 49 (2010).\\
21. S. Sazonov, M. Revnivtsev, A. Lutovinov, R. Sunyaev, and S. Grebenev, \aap\ {\bf 421}, L21 (2004).\\
22. D.J. Schlegel, D.P. Finkbeiner, and M. Davis, \apj\ {\bf 500}, 525 (1998).\\
23. P. Ubertini, F. Lebrun, G. Di Cocco, A. Bazzano, A.J. Bird, K. Broenstad,  A. Goldwurm, G. La Rosa, et al., \aap\ {\bf 411}, L131 (2003).\\
24. A. Udalski, \apj\ {\bf 590}, 284 (2003).\\
25. W. Wegner, MNRAS {\bf 374}, 1549 (2007).\\
26. W. Wegner, Acta Astronomica {\bf 64}, 261 (2014).\\
27. C. Winkler, T.J.-L. Courvoisier, G. Di Cocco, N. Gehrels, A. Gimenez,    S. Grebenev, W. Hermsen, J.M. Mas-Hesse, F. Lebrun, et al.,  \aap\ {\bf 411}, L1 (2003).\\
28. L. Yungelson, M. Livio, A. Tutukov, and S.J. Kenyon), \apj\ {\bf 447}, 656 (1995).\\

\end{document}